\documentclass[12pt]{article}

\usepackage{graphicx}
\usepackage{amsmath,amsfonts}
\usepackage{epstopdf}
\usepackage[top=2in, bottom=1.5in, left=1in, right=1in]{geometry}
\usepackage{tensor}
\usepackage{slashed}
\usepackage{cite}

\allowdisplaybreaks

\def \be {\begin{equation}}
\def \ee {\end{equation}}
\def \bea {\begin{eqnarray}}
\def \eea {\end{eqnarray}}

\newcommand{\bpsi}{\overline{\psi}}

\newcommand{\softs}{\underline{s}}

\newcommand{\hp}{\hat{p}}
\newcommand{\hk}{\hat{k}}

\newcommand{\qslash}{\slashed{q}}
\newcommand{\pslash}{\slashed{p}}
\newcommand{\kslash}{\slashed{k}}
\newcommand{\sslash}{\slashed{s}}

\title{Soft Photon Theorem for High Energy Amplitudes in Yukawa and Scalar Theories}
\author{Hualong Gervais\\ C.N.\ Yang Institute for Theoretical Physics\\ and Department of Physics and Astronomy\\
Stony Brook University, Stony Brook, NY 11794-3840 USA}

\begin{document}

\begin{flushright}
YITP-SB-17-11
 \end{flushright}

{\let\newpage\relax\maketitle}

\begin{abstract}
We study the emission of soft photons coupling to high energy fixed angle scattering processes at first order in the electromagnetic coupling but to all loop orders in a class of theories without soft divergences, including massive and massless Yukawa and scalar theories. We adapt a method introduced by del Duca for quantum electrodynamics to show that subleading corrections to the soft photon theorem are sensitive to the structure of non leading external jets of collinear lines. Our techniques are based on a power counting analysis of loop integrals and an application of jet Ward identities. We also apply Grammer and Yennie's decomposition to isolate separately gauge invariant contributions to the soft photon expansion. These are interpreted as infrared sensitive matrix elements coupling to a field strength tensor.
\end{abstract}

\tableofcontents




\section{Introduction} \label{sec:introduction}

The subject of the emission of soft particles in quantum field theory has a long history dating back to the classic theorems of Low, Burnett and Kroll, and Weinberg  \cite{Low:1958sn, Burnett:1967km,Weinberg:1965nx}. The leading term in the soft particle energy $q^0$ universally behaves as $1/q^0$ and comes from dressing an external line with a tree level vertex. The form of a soft theorem is strongly influenced by the underlying symmetries of the theory. In QED, Low's classic result shows that the next to leading term is fixed by gauge invariance, implemented through the use of Ward identities. More recently, it has been shown that soft graviton theorems can be realized as the Ward identities of a symmetry of the gravitational S-matrix \cite{He:2014laa,Strominger:2013jfa}. This observation is part of a renewed interest in soft theorems in both gauge and gravitational theories. In particular, Ref. \cite{Cachazo:2014fwa} has used the BCFW relation \cite{Britto:2004ap,Britto:2005fq} to not only derive Weinberg's leading term for soft graviton emission but also determine the next and next to next to leading terms at tree level. It was later confirmed that these higher order terms in the soft graviton theorem can be understood from the gravitational Ward identity \cite{Weinberg:1964ew} following a treatment similar to Low's analysis \cite{Bern:2014vva}. Loop corrections were investigated in \cite{Bern:2014oka,He:2014bga,Cachazo:2014dia}.

In the case of gauge theories, the resummation of logarithms associated with soft and collinear gluon emissions has been applied to numerous collider observables \cite{Sterman:1986aj,Catani:1989ne,Korchemsky:1993uz,Korchemsky:1992xv,Forte:2002ni,Contopanagos:1996nh,Banfi:2004yd,Becher:2006nr,Luisoni:2015xha} and  loop corrections to the subleading term in soft theorems are important for applications to precision studies of the Standard Model \cite{Bonocore:2015esa,Bonocore:2016awd}.

Soft theorems in gravity and gauge theories have also been studied from several other viewpoints, including scattering equations \cite{Cachazo:2013hca,Cachazo:2013iea,Schwab:2014xua}, string theory techniques \cite{Geyer:2014lca, Schwab:2014fia, Bianchi:2014gla}, path integral and diagrammatic methods \cite{White:2009aw,White:2011vh, White:2011yy,White:2013ye,White:2014qia,   Laenen:2008gt,Laenen:2010uz,Luna:2016idw}, and effective field theory \cite{Larkoski:2014bxa}. In particular, Ref. \cite{Larkoski:2014bxa}  stresses the importance of matrix elements involving higher dimension operators, which we will derive from an independent point of view.

In QED, loop corrections were studied long ago by del Duca, who recognized the importance of carefully handling infrared logarithms \cite{DelDuca:1990gz}. Given $m$ the mass of the external particles and $E$ the center of mass energy, del Duca showed that in the high energy limit, the original form of Low's theorem only holds in the small region $q^0 \ll m^2 / E$. When $m^2/E < q^0 < m$, loop corrections to the first subleading term appear. These corrections depend on the structure of jets of lines collinear to the external particles. 

Our purpose in this paper is to revisit del Duca's analysis by considering soft photon emission in the regime where $ q \sim m^2 / E$.  We couple the electromagnetic field to massive fermions interacting with massless scalars through Yukawa interactions in four dimensions. The scalars are allowed to interact via a quartic potential but will be kept neutral for simplicity. Our final form of the soft photon theorem considers the case where all external hard particles are outgoing fermions or antifermions. The extension of our results to the cases where the scalars are charged and allowed to appear as external particles is straightforward but involves more terms in our final formulas. Our results will hold to first order in the electric charge but to all orders in the Yukawa and $\phi^4$ couplings.

Yukawa and $\phi^4$ theories provide a non trivial testing ground for our ideas. We will show that not only is soft photon emission sensitive to collinear jets, but also that the class of jets contributing to the amplitude is broader than those corresponding to the leading contribution in the elastic amplitude. The term ``elastic'' here and below refers only to the absence of energy loss to the electromagnetic field.

This paper is organized as follows. Sections \ref{sec:review_low_theorem} and \ref{sec:linear_expansion_failure} review the basic reasoning of Low's theorem and identify the need for further study of the analytic structure of loop amplitudes at high energies. In reviewing Low's theorem, we take this opportunity to discuss the transition between radiative and elastic kinematics.  In Secs.\ \ref{sec:power_counting} and \ref{sec:power_counting_applied}, we briefly review the power counting techniques we have employed to find all pinch surfaces contributing to the radiative amplitude up to order $O(q^0)$. In Sec.\ \ref{sec:factorization}, we define a factorized elastic amplitude incorporating contributions from all pinch surfaces relevant to the soft photon theorem at high energies. Sections \ref{sec:example_fs} and \ref{sec:example_soft}  present explicit examples that illustrate our general results based on power counting and also exhibit the non analytic contributions in loop amplitudes described in \ref{sec:linear_expansion_failure}. In Sec.\ \ref{sec:low_argument} , we carry out Low's argument, adapted to the factorized amplitudes that we identify in Sec.\ \ref{sec:factorization}. This will result in a preliminary form of the expansion of the radiative amplitude, which we further refine in Sec.\ \ref{sec:kg_decomposition} using a decomposition into photon polarizations inspired from Grammer and Yennie's work \cite{Grammer:1973db}.




\section{Soft photon emission at high energies} \label{sec:high_energy_photon}

In the high energy regime where $q \sim m^2 /E$ and $E \gg m$, we will see that a complete treatment of Low's theorem requires a study of the analytic structure of loop integrals. To introduce the need for this analysis, we begin by reviewing Low's classic approach to soft radiation in the following subsection. The original treatment of Low appears in \cite{Low:1958sn} and his analysis was also adapted to non-abelian gauge theory and gravity in \cite{Bern:2014vva}. While reviewing Low's theorem, we will discuss the issue of retaining momentum conservation when transitioning between the kinematics with and without an external photon. This point is often neglected in the literature, but we believe it is relevant if one is to apply Low's theorem to realistic scenarios.

Low's theorem is traditionally stated as the expansion of a radiative amplitude $\mathcal{M}(q)$ in powers of $q$ up to order $q^0$,
\begin{align} \label{eq:classic_low}
\mathcal{M}(q) &= \frac{1}{q}\sigma_{-1}  + \sigma_0 \, ,
\end{align}
where the coefficients $\sigma_{-1}$ and $\sigma_0$ are built from the quantities at hand in the problem, such as $m$ and $E$. However, since we are considering the regime where $q \sim m^2/E$, the soft momentum $q$ is no longer the only ``small'' quantity in the problem. Consequently, a quantity scaling as $qE/m^2$ would be of the same order as the first subleading term $\sigma_0$.

To get a complete soft photon expansion, it is important to identify carefully all contributions to the radiative amplitude that are of the same order of magnitude as the leading and subleading terms in Eq.\ \eqref{eq:classic_low}. This requires us to define a common ``small'' scale in terms of which all orders of magnitude will be expressed. With that objective in mind, we treat the total center of mass energy $E$ as the scale of hard processes. We are interested in the regime where the dimensionless parameter $\lambda \equiv \frac{m}{E}$ is much less than $1$ and use it to quantify what we mean by ``small''. Using this notation, we then have that $q\sim \lambda^2 E$ and $m \sim \lambda E$. Further, given an arbitrary quantity $a$, the notation $a = O(\lambda^\gamma)$ will mean that there exists some constant $A$ such that $|a| \leq A \lambda^\gamma$. The constant $A$ can be constructed with the appropriate power of $E$ to have the same dimension as $a$ but may not depend on $q$ or $m$. For instance, we have $q = O(\lambda^2)$ and $m = O(\lambda)$. Low's theorem is then an expansion going from $O(\lambda^{-2})$ to $O(\lambda^0)$.  With these definitions established, we proceed to reviewing Low's theorem.

\subsection{Review of Low's theorem} \label{sec:review_low_theorem}

It is enough to consider Low's original case of scalar Compton scattering to illustrate our points on the importance of the infrared behavior of loop integrals. Therefore, consider a radiative Compton scattering process
\begin{align}
f(p_1) + s(k_1) \rightarrow f(p_2) + s(k_2) + \gamma(q) \, ,
\end{align}
where a charged fermion and a neutral scalar of respective momenta $p_1$ and $k_1$ scatter into a fermion and scalar of momenta $p_2$ and $k_2$ while a soft photon of momentum $q$ is emitted. As defined in the Introduction, the corresponding elastic amplitude is the same process without the emission of the soft photon. Naturally, the momenta $p_1$, $p_2$, $k_1$, and $k_2$ are all on-shell and 
\begin{align}
p_1 + k_1 &= p_2 + k_2 + q \, ,
\end{align}
as required by momentum conservation.

\begin{figure}[h!]
\centering
\includegraphics[width = 0.75 \textwidth]{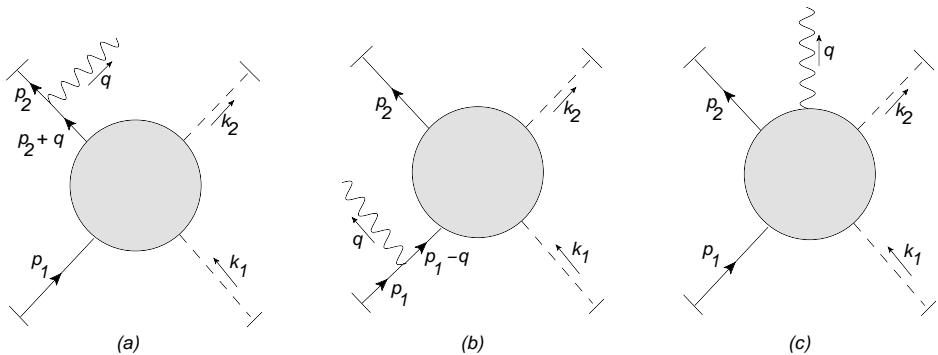}
\caption{Diagrams (a) and (b) represent the external amplitudes. Diagram (c) is the internal amplitude where the soft photon is attached to an internal fermion propagator. }
\label{fig:compton_ext_int}
\end{figure}

The Feynman diagrams contributing to the radiative amplitude are generated by attaching an external soft photon line to the diagrams contributing to the elastic amplitude. We distinguish between two types of radiative emission amplitudes. Those where the soft photon is attached to an external fermion line are called ``external'' radiative amplitudes while those where the soft photon is attached to an internal fermion line are called ``internal'' radiative amplitudes. These two types of amplitudes are illustrated in Fig.\ \ref{fig:compton_ext_int}. In Sec.\ \ref{sec:low_argument}, we will adapt these definitions to a factorized form of the radiative amplitude.


\begin{figure}[h!]
\centering
\includegraphics[width = 0.75 \textwidth]{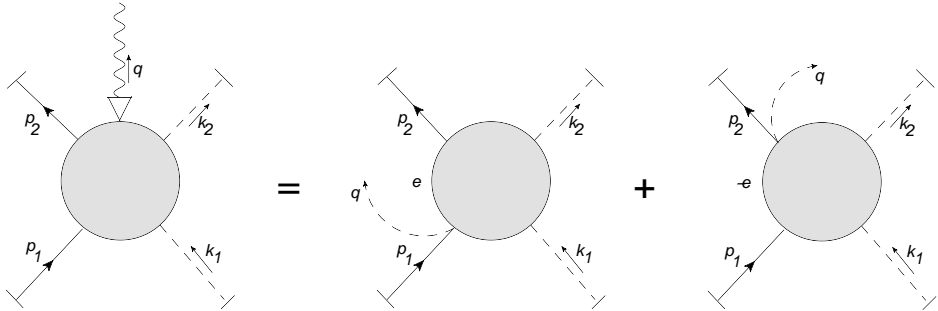}
\caption{The Ward identity relates the internal radiative amplitude to the elastic amplitude with external momenta shifted by $q$. The arrow at the end of the photon line on the left hand side indicates that the current operator corresponding to the emitted photon is contracted with the photon momentum $q$. On the right, the photon momentum $q$ is pictured as exiting the diagram through a composite scalar-fermion-photon vertex.}
\label{fig:compton_ward_identity}
\end{figure}

The explicit forms of the external radiative amplitudes from Fig.\ \ref{fig:compton_ext_int} are
\begin{align}\label{eq:compton_external_explicit}
M^{ext,\mu}_{a}(p_1, p_2, k_1, k_2, q) &= \bar{u}(p_2) (-i e \gamma^\mu) \frac{i}{\pslash_2 + \qslash - m } \tilde{M}_{el}(p_1, p_2 + q, k_1, k_2) u(p_1) \, , \nonumber \\
M^{ext,\mu}_{b}(p_1, p_2, k_1, k_2, q) &= \bar{u}(p_2) \tilde{M}_{el}(p_1 - q, p_2,  k_1, k_2) \frac{i}{\pslash_1 - \qslash - m}(-ie \gamma^\mu) u(p_1) \, .
\end{align}
In the above, we denote the elastic amplitude stripped of the external spinors by $\tilde{M}_{el}(\dots)$. The internal radiative amplitude is denoted with the symbol $M^{int,\mu}(p_1, p_2, k_1, k_2, q)$. This notation allows us to express the QED Ward identity in a form that corresponds to the separation of the photon emission amplitude into emission from external or internal legs,
\begin{align} \label{eq:low_insight}
q_\mu M^{ext,\mu}_{a} + q_\mu M^{ext,\mu}_{b} + q_\mu M^{int,\mu} = 0 \, .
\end{align}
Substituting the explicit forms for the external amplitudes of Eq.\ \eqref{eq:compton_external_explicit} into Eq.\ \eqref{eq:low_insight}, we find that the Ward identity becomes
\begin{align}\label{eq:low_insight_explicit}
e\, \bar{u}(p_2) \tilde{M}_{el}(p_1, p_2 + q, k_1, k_2) u(p_1) + (-e)\, \bar{u}(p_2) \tilde{M}_{el}(p_1 - q , p_2, k_1, k_2) u(p_1) + q_\mu M^{int,\mu} = 0\, .
\end{align}
This Ward identity is illustrated in Fig. \ref{fig:compton_ward_identity}.

Equation \eqref{eq:low_insight_explicit} allows us to solve for the internal amplitude $M^{int,\mu}$ in terms of derivatives of the elastic amplitude. We assume, following Low, that it is consistent to expand the stripped elastic amplitude $\tilde{M}_{el}(\dots)$ in powers of $q$ about the point $(p_1, k_1, p_2, k_2)$. Using charge conservation, one then finds
\begin{align}
q_\mu M^{int, \mu}\, = & -e\,q_\mu \,  \bar{u}(p_2)  \frac{\partial}{\partial p_{1,\mu}} \tilde{M}_{el}(p_1, p_2, k_1, k_2)\, u(p_1) \nonumber \\
& -e\, q_\mu \,  \bar{u}(p_2) \frac{\partial}{\partial p_{2,\mu}} \tilde{M}_{el}(p_1, p_2, k_1, k_2)\, u(p_1)  + O(\lambda^4) \, .
\end{align}
A possible problem with this procedure, as discussed by Burnett and Kroll in \cite{Burnett:1967km}, is that we are left with a formula where the elastic amplitude is evaluated at a point outside the locus of momentum conservation. This unphysical formula could be ambiguous in realistic applications and thus an alternative would be preferrable.

\begin{figure}[h!]
\centering
\includegraphics[width = 0.25 \textwidth]{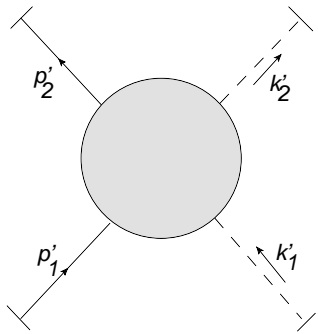}
\caption{The elastic $fs \rightarrow \, fs$ amplitude involves no photon emission. The external momenta are on-shell and obey momentum conservation.}
\label{fig:compton_elastic}
\end{figure}

Following Burnett and Kroll, we define an elastic momentum configuration $\{p_1^\prime, k_1^\prime, p_2^\prime, k_2^\prime \}$ where $p_1^\prime$ and $k_1^\prime$ are the respective momenta of the incoming fermion and scalar while $p_2^\prime$ and $k_2^\prime$ are the momenta of the outgoing fermion and scalar -- see Fig. \ref{fig:compton_elastic}. These elastic momenta are all on-shell and obey momentum conservation in the absence of the soft photon,
\begin{align}
p_1^\prime + k_1^\prime &= p_2^\prime + k_2^\prime \, .
\end{align}
We want the elastic momenta to be shifted slightly away from the radiative configuration. Hence, we introduce the small deviations $\xi_1(q)$, $\xi_2(q)$, $\eta_1(q)$, and $\eta_2(q)$ satisfying
\begin{align}
p_i &= p_i^\prime(q) + \xi_i(q)\quad \text{for $i = 1,2$,} \nonumber \\
k_i &= k_i^\prime(q) + \eta_i(q) \quad \text{for $i = 1,2$.}
\end{align}
We also want that when $q=0$, the radiative fermion and scalar momenta coincide with the elastic ones. This motivates the requirement that the $\xi_i$'s and $\eta_i$'s be polynomials in $q$ whose leading term is linear in $q$, and in particular, $\xi_i, \eta_i = O(\lambda^2)$, as for $q^\mu$.


Now, insisting on preserving momentum conservation and having on-shell particles in the elastic amplitude induces the following constraints on the $\xi_i$'s and $\eta_i$'s,
\begin{align}  \label{eq:xi_eta_constraints}
\xi_1 + \eta_1 - \xi_2 - \eta_2 &= q \, , \nonumber \\
2 p_i \cdot \xi_i &= \xi_i^2 \quad \text{for $i=1,2$,}\nonumber \\
2 k_i \cdot \eta_i &= \eta_i^2 \quad \text{for $i=1,2$. }
\end{align}
To leading order in $\lambda$, the latter two equations in \eqref{eq:xi_eta_constraints} above become
\begin{align} \label{eq:xi_eta_dot_pk}
p_i \cdot \xi_i &=  k_i \cdot \eta_i = 0 \quad \text{for $i = 1,2$.}
\end{align}

To construct the $\xi_i$'s and $\eta_i$'s at leading order in $\lambda$, it is convenient to introduce an orthonormal frame about each external fermion and scalar momentum. About the fermion momenta $\vec{p}_1$ and $\vec{p}_2$, we introduce the three dimensional vectors $\vec{n}_{p_i}$, $\vec{\bar{n}}_{p_i}$, and $\vec{e}_{p_i}$. The vector $\vec{e}_{p_i} = \vec{p}_i / |\vec{p}_i|$ points in the direction parallel to $\vec{p}_i$ while the vectors $\vec{n}_{p_i}$ and $\vec{\bar{n}}_{p_i}$ span the plane orthonormal to $\vec{p}_i$. Likewise, to each scalar's momentum $\vec{k}_i$, we associate the orthonormal frame $\vec{n}_{k_i}$, $\vec{\bar{n}}_{k_i}$, and $\vec{e}_{k_i}$. We then decompose the $\vec{\xi}_i$'s and $\vec{\eta}_i$'s in their corresponding orthonormal bases
\begin{align}\label{eq:xi_eta_decomposition}
\vec{\xi}_i &= \alpha_{p_i} \vec{n}_{p_i} + \beta_{p_i} \vec{\bar{n}}_{p_i} + \delta_{p_i} \vec{e}_{p_i} \quad \text{for $i =1,2$,} \nonumber \\
\vec{\eta}_i &= \alpha_{k_i} \vec{n}_{k_i} + \beta_{k_i} \vec{\bar{n}}_{k_i} + \delta_{k_i} \vec{e}_{k_i} \quad \text{for $i =1,2$,}
\end{align}
where all coefficients $\alpha_*$, $\beta_*$, and $\delta_*$ are $O(\lambda^2)$.

Since the $p_i$'s and $k_i$'s are on-shell, we have
\begin{align}
p_i^0 &= \sqrt{|\vec{p}_i|^2 + m^2} \quad \text{for $i=1,2$, }\nonumber \\
k_i^0 &= |\vec{k}_i| \quad\quad \quad\quad \quad\text{for $i=1,2$. }
\end{align}
Combining this with \eqref{eq:xi_eta_dot_pk} and \eqref{eq:xi_eta_decomposition}, we find that
\begin{align}
\xi_i^0 &= \delta_{p_i} v_{p_i} \quad \text{for $i=1,2$,} \nonumber \\
\eta_i^0 &= \delta_{k_i}\,\quad \quad \text{for $i= 1,2$,}
\end{align}
where we have  defined the velocities $v_{p_i} \equiv |\vec{p}_i| / \sqrt{|\vec{p}_i|^2 + m^2} $ for $i=1,2$. This fully characterizes the components $\xi_i^0$'s and $\eta_i^0$'s, and takes care of ensuring that the elastic momenta are on-shell to leading order in $\lambda$. We still need to solve for momentum conservation in Eqs. \eqref{eq:xi_eta_constraints}. This requirement is most conveniently written in matrix notation,
\begin{align}\label{eq:linear_system_xi_eta}
\begin{pmatrix}
0 & 0 & v_{p_1}  & 0 & 0 & 1  & 0 & 0 & -v_{p_2} & 0 & 0 & -1 \\
\vec{n}_{p_1} & \vec{\bar{n}}_{p_1} & \vec{e}_{p_1} & \vec{n}_{k_1} & \vec{\bar{n}}_{k_1} & \vec{e}_{k_1} &  -\vec{n}_{p_2} & -\vec{\bar{n}}_{p_2} & -\vec{e}_{p_2} &  -\vec{n}_{k_2} & -\vec{\bar{n}}_{k_2} & -\vec{e}_{k_2} 
\end{pmatrix}
\begin{pmatrix}
\alpha_{p_1}\\
\beta_{p_1}\\
\delta_{p_1}\\
\alpha_{k_1}\\
\beta_{k_1}\\
\delta_{k_1}\\
\alpha_{p_2}\\
\beta_{p_2}\\
\delta_{p_2}\\
\alpha_{k_2}\\
\beta_{k_2}\\
\delta_{k_2}
\end{pmatrix}
&=
\begin{pmatrix}
q^0 \\
\vec{q}
\end{pmatrix}\, .
\end{align}
The columns of the leftmost matrix above are made of the components of the orthonormal frame vectors relative to some fixed common frame. Assuming the rank of the resulting matrix to be maximal, we can determine the four coefficients $\alpha_{p_1}$, $\beta_{p_1}$, $\delta_{p_1}$, and $\alpha_{k_1}$ in terms of the remaining $\alpha_*$, $\beta_*$, $\delta_*$, and $q$. If we choose the undetermined components to be $O(\lambda^2)$, then the solution for $\alpha_{p_1}$, $\beta_{p_1}$, $\delta_{p_1}$, and $\alpha_{k_1}$ will also be $O(\lambda^2)$. Therefore, the construction we have outlined allows us to derive $\xi_i$'s and $\eta_i$'s that are of the same order of magnitude as the soft momentum $q$. It is then clear that one must include these deviations from the elastic configuration in a complete expansion of the radiative amplitude.

Returning to Eq.\ \eqref{eq:low_insight_explicit}, we proceed with the expansion of $\tilde{M}_{el}(\dots)$ to $O(\lambda^2)$ about $(p_1^\prime, p_2^\prime, k_1^\prime, k_2^\prime)$, taking the $\xi_i$'s and $\eta_i$'s into account,
\begin{align} \label{eq:low_insight_explicit_expanded}
q_\mu & M^{int,\mu}(p_1, p_2, k_1, k_2, q) = \nonumber \\
& e \, \bar{u}(p_2)\left(\tilde{M}_{el}(p_1^\prime,\dots) + \sum_{i=1,2} \left( \xi_i^\mu \frac{\partial}{\partial p_i^\mu}+ \eta_i^\mu \frac{\partial}{\partial k_i^\mu}\right) \tilde{M}_{el}(p_1^\prime,\dots) - q^\mu \frac{\partial}{\partial p_1^\mu} \tilde{M}_{el}(p_1^\prime,\dots) \right)u(p_1) \nonumber \\
& -e \, \bar{u}(p_2)\left(\tilde{M}_{el}(p_1^\prime,\dots) + \sum_{i=1,2} \left( \xi_i^\mu \frac{\partial}{\partial p_i^\mu}+ \eta_i^\mu \frac{\partial}{\partial k_i^\mu}\right) \tilde{M}_{el}(p_1^\prime,\dots)+ q^\mu \frac{\partial}{\partial p_2^\mu} \tilde{M}_{el}(p_1^\prime,\dots) \right)u(p_1) \nonumber \\
&+O(\lambda^4) \, .
\end{align}
When taking derivatives in the above, we treat the momenta $p_1$, $p_2$, $k_1$, and $k_2$ as variables upon which the elastic amplitude $\tilde{M}_{el}$ depends. The differentiated amplitude is then evaluated at the elastic configuration $(p_1^\prime, p_2^\prime, k_1^\prime, k_2^\prime)$. Charge conservation reduces \eqref{eq:low_insight_explicit_expanded} to
\begin{align}\label{eq:equation_for_internal}
q_\mu  M^{int,\mu}(p_1, p_2, k_1, k_2, q) =& -e \, q^\mu \,  \bar{u}(p_2)\frac{\partial}{\partial p_1^\mu} \tilde{M}_{el}(p_1^\prime, p_2^\prime, k_1^\prime, k_2^\prime)\, u(p_1) \nonumber \\
&  -e \, q^\mu \, \bar{u}(p_2)  \frac{\partial}{\partial p_2^\mu} \tilde{M}_{el}(p_1^\prime,p_2^\prime, k_1^\prime, k_2^\prime)\, u(p_1) + O(\lambda^4) \, .
\end{align}
We see that the $\xi_i$ and $\eta_i$ terms have cancelled because of charge conservation. This means that the way we choose to transition from radiative to elastic kinematics does not affect the essence of Low's theorem, which is the determination of the internal radiative amplitude in terms of the external one. To obtain a full expansion of the radiative amplitude however, we also need to expand the external amplitudes in Eq. \eqref{eq:compton_external_explicit} to $O(\lambda^2)$. As we shall see, the $\xi_i$ and $\eta_i$ terms no longer cancel in the expansion of these external amplitudes.

Note that the elastic configuration $p_1^\prime, p_2^\prime, k_1^\prime, k_2^\prime$ is not unique. If we were to solve Eq. \eqref{eq:linear_system_xi_eta} by making a different choice for the undetermined coefficients, we would obtain a different set of $\xi_i$'s and $\eta_i$'s. However, as long as we choose coefficients that are $O(\lambda^2)$, the difference in the elastic configuration we obtain will also be $O(\lambda^2)$. This would induce corrections to the internal radiative amplitude that are beyond $O(\lambda^0)$, and hence beyond our order of accuracy in Low's theorem.

The obvious particular solution to Eq. \eqref{eq:equation_for_internal} is derived by simply ``factoring out'' the soft photon momentum $q_\mu$,
\begin{align}\label{eq:low_classic_result}
M^{int, \mu}(p_1, p_2, k_1, k_2, q) = & -e \,  \bar{u}(p_2)  \frac{\partial}{\partial p_{1,\mu}} \tilde{M}_{el}(p_1^\prime, p_2^\prime, k_1^\prime, k_2^\prime)\, u(p_1) \nonumber \\
& -e \,  \bar{u}(p_2) \frac{\partial}{\partial p_{2,\mu}} \tilde{M}_{el}(p_1^\prime, p_2^\prime, k_1^\prime, k_2^\prime)\, u(p_1)  + O(\lambda^2) \, .
\end{align}
However, one may inquire about the possibility of separately gauge invariant contributions to $M^{int,\mu}$. These have the generic form
\begin{align} \label{eq:gauge_invariant_terms}
B^\mu(l,q) = & \sum_{l \in S} f_{1,l} (S,q)\, (l\cdot q \,  q^\mu  - q^2 l^\mu ) \,  \bar{u}(p_2) u(p_1) \nonumber \\
&+ \sum_{l \in S}  f_{2,l} (S,q)\, (l\cdot q \,  q^\mu  - q^2 l^\mu) \,  \bar{u}(p_2) \gamma_5 u(p_1) \nonumber \\
&+ f_3(S, q) \,  \bar{u}(p_2) [\gamma^\mu, \qslash]  u(p_1) \, ,
\end{align}
where $S = \{ p_1, p_2, k_1, k_2 \}$ is the set of external momenta excluding the soft photon momentum. The tensor structures corresponding to $f_1$, $f_2$, and $f_3$ are of order $\lambda^4$, $\lambda^4$, and $\lambda^2$ respectively. Accordingly, to have contributions of order $\lambda^0$, $f_1$, $f_2$, and $f_3$ must have enhancements of orders $\lambda^{-4}$, $\lambda^{-4}$, and $\lambda^{-2}$. Such enhancements do not appear in the low energy regime $E \sim m$ where the classic form of Low's theorem holds. At high energies $E \gg m$ however, these enhancements are intimately linked to the infrared behavior of loop integrals and the need to generalize the classic form of Low's theorem, as we will see in Secs. \ref{sec:linear_expansion_failure} and \ref{sec:example_fs}.

We conclude this section by showing the full radiative amplitude deduced from the classic form of Low's argument,
\begin{align}
M^\mu = \,  & \bar{u}(p_2)(-ie \gamma^\mu) \frac{i}{\pslash_2 + \qslash - m}\tilde{M}_{el}(p_1^\prime, p_2^\prime,  k_1^\prime, k_2^\prime) \,  u(p_1) \nonumber \\
&  + \bar{u}(p_2) \tilde{M}_{el}(p_1^\prime, p_2^\prime, k_1^\prime, k_2^\prime) \frac{i}{\pslash_1 - \qslash - m}(-ie \gamma^\mu)  \, u(p_1) \nonumber \\
&  + \bar{u}(p_2) \, \left[ \sum_{i=1,2} \left( \xi_i^\alpha \frac{\partial}{\partial p_i^\alpha} + \eta_i^\alpha \frac{\partial}{\partial k_i^\alpha}\right) - q^\alpha \frac{\partial}{\partial p_1^\alpha}  \right] \tilde{M}_{el}(p_1^\prime , p_2^\prime, k_1^\prime, k_2^\prime) \frac{i}{\pslash_1 - \qslash - m}(-ie \gamma^\mu)\,  u(p_1) \nonumber \\
& + \bar{u}(p_2) (-i e \gamma^\mu) \frac{i}{\pslash_2+\qslash - m } \,  \left[ \sum_{i=1,2} \left( \xi_i^\alpha \frac{\partial}{\partial p_i^\alpha} + \eta_i^\alpha \frac{\partial}{\partial k_i^\alpha} \right) + q^\alpha \frac{\partial}{\partial p_2^\alpha}  \right] \tilde{M}_{el}(p_1^\prime, p_2^\prime,  k_1^\prime, k_2^\prime) \, u(p_1) \nonumber \\
 & -e \,  \bar{u}(p_2)  \frac{\partial}{\partial p_{1,\mu}} \tilde{M}_{el}(p_1^\prime, p_2^\prime, k_1^\prime, k_2^\prime)\, u(p_1) \nonumber \\
& -e \,  \bar{u}(p_2) \frac{\partial}{\partial p_{2,\mu}} \tilde{M}_{el}(p_1^\prime, p_2^\prime, k_1^\prime, k_2^\prime)\, u(p_1)  \nonumber \\
& + O(\lambda^2) \, ,
\end{align}
where we retain the necessary $\xi_i$ and $\eta_i$ dependence for the external connections, which enter with different Dirac structure, and hence do not cancel in general. In the next section, we show precisely why the classic form of Low's theorem we have just described requires generalization in the regime $q = O(\lambda^2)$.

\subsection{Failure of the linear expansion}\label{sec:linear_expansion_failure}

In Sec.\ \ref{sec:review_low_theorem}, our ability to deduce $M^{int,\mu}$ from the Ward identity \eqref{eq:low_insight} depends on being able to expand the elastic amplitude $\tilde{M}_{el}(\dots)$ in \eqref{eq:low_insight_explicit} to linear order in $q$. The accuracy of this expansion is intimately tied to the infrared behavior of the loop integrals contributing to the radiative and elastic amplitude. 

Once all loop integrals contributing to the radiative amplitude have been carried out, some regions of the loop integration will have yielded functions of $q$ that are either singular at $q = 0$ or are analytic with a ``very small'' radius of convergence. Examples of the former include pole terms such as $1/p_i \cdot q$. The latter category includes logarithmic terms such as $f(q) \equiv \log(1 + a p_i \cdot q / m^2)$ where $a = O(1)$ is some constant. The Taylor series for this logarithmic function of $q$ has radius of convergence $R \sim m^2/E = O(\lambda^2)$ since there is a branch cut within a distance of order $O(\lambda^2)$ of $q=0$. When expanding $f(q)$ to linear order, the remainder has a small upper bound only in the vanishingly small region $q << m^2/E$. It is convenient here to apply the term ``non analytic'' to functions whose power series have radius of convergence $R = O(\lambda^2)$. Our claim is then that in general, loop integrals have singular and non analytic contributions that either cannot be Taylor expanded about $q=0$ altogether, or whose linear expansion in $q$ is accurate only for vanishingly small photon momenta $q << O(\lambda^2)$. Either way, in our region of interest, such contributions prevent us from carrying out Low's argument as described in Sec. \ref{sec:review_low_theorem}. 

To extend Low's theorem to high energy scattering, it is necessary first to identify singular and non analytic contributions, and then factorize them from terms that can be legitimately expanded to linear order in $q$. Fortunately, identifying these contributions can be done by studying the loop integrand rather than fully evaluating the loop integral \cite{Landau:1959fi,Coleman:1965xm,Libby:1978qf,Collins:1989gx,Sterman:1995fz,Sterman:1994ce,Collins:2011zzd}. We introduce these methods through an example.

Consider a triangle integral where two massive on-shell particles of mass $m$ and momenta $p_1$ and $p_2$ exchange a massless scalar,
\begin{align} \label{eq:power_counting_example}
I \equiv \int d^4k \frac{1}{(k^2 + i\epsilon) ( (k+p_1)^2 - m^2 + i \epsilon) ( (k+p_2)^2 - m^2 + i\epsilon)} \, ,
\end{align}
with $p_1^2 = p_2^2 = m^2$ and $(p_1-p_2)^2 < 0$. Although they would have to be included in general, we ignore numerator factors in this illustrative example. Our goal is to locate regions of the $d^4k$ integration that may result in singular or non analytic terms. At high energy, the coordinates best suited to this goal are light cone coordinates. An arbitrary vector $v$ is defined by its components $(v^+, v^-, v_T)$ with the standard definitions
\begin{align}
v^\pm &\equiv \frac{1}{\sqrt{2}}(v^0\pm v^3) \, ,  \nonumber \\
v_T &\equiv (v^1, v^2) \, .
\end{align}
Scalar products take the form
\begin{align}
v\cdot w = v^+ w^- + v^- w^+ - v_T \cdot w_T \, .
\end{align}
Collinear and soft momenta are defined by the scaling of their light cone coordinates. Suppose that $p_1$ is moving in the $z$ direction. The components of a momentum $k$ collinear to $p_1$ scale as $(1, \lambda^2, \lambda)E$. Those of a soft momentum, on the other hand, scale as $(\lambda^2, \lambda^2, \lambda^2)E$ .   An arbitrary hard momentum scales as $(1,1,1)E$. Note that in our example, $p_2$ is hard relative to $p_1$. Focusing on the region where $k$ is collinear to $p_1$ in our example, it is straightforward to see that
\begin{align} \label{eq:basic_scaling}
k^2 &= O(\lambda^2) \, ,\nonumber \\
k^2 + 2 p_1 \cdot k &= O(\lambda^2) \, , \nonumber \\
k^2 + 2 p_2 \cdot k &= 2 p_2^- k^+ + O(\lambda) = O(\lambda^0)\, .
\end{align}

We can now explain why it is inaccurate to expand $\tilde{M}_{el}(p_1 - q, p_2, k_1, k_2)$ to linear order in $q$ by studying the integrand. Consider Eq. \eqref{eq:power_counting_example} with the external momentum $p_1$ replaced with $p_1 - q$ similarly to $\tilde{M}_{el}(p_1 - q, p_2, k_1, k_2)$ in \eqref{eq:low_insight_explicit},
\begin{align}
I^\prime \equiv \int d^4k \frac{1}{(k^2 + i\epsilon) (k^2+2 (p_1 - q) \cdot k + q^2 - 2 p_1 \cdot q  + i\epsilon) ( k^2 + 2 p_2 \cdot k + i\epsilon)} \, .
\end{align}
The sum of the invariants with a factor of $q$ in the middle denominator is
\begin{align*}
q^2 -2 p_1 \cdot q - 2 k \cdot q &= O(\lambda^2) \, ,
\end{align*}
 which is of the same order of magnitude as $k^2 + 2 p_1 \cdot k$ in region \eqref{eq:basic_scaling}. Therefore, when $q$ flows through a momentum collinear to $p_1$, we may not treat terms in propagator denominators with factors of $q$ as small quantities in which we can expand using a power series. Further, the scaling $k^2 + 2 p_1 \cdot k = O(\lambda^2)$  implies that $k+p_1$ is very close to the mass shell. Similar conclusions apply when $k$ is collinear to $p_2$ rather than $p_1$.

It turns out that internal momenta going on-shell are a necessary condition for having singular or non analytic terms. To see this, consider the generic multiloop integral
\begin{align}\label{eq:generic_loop_integral}
\tilde{F}(p_1, \dots, p_n) &= \int \prod_{l=1}^L d^4 k_l \int \prod_{j=1}^M d\alpha_j \, \delta\left(1 - \sum_{j=1}^M \alpha_j\right) \frac{\mathcal{N}(\{p_i, k_l, \alpha_j\})}{\left[\sum_{j=1}^M \alpha_j (l_j^2(\{p_i, k_l\}) - m_j^2)+i\epsilon \right]^{\sum_{j=1}^M\delta_j}} \, ,
\end{align}
where the Feynman parameters $\alpha_1, \dots, \alpha_M$ have been introduced. The numerator factor $\mathcal{N}(\{p_i, k_l, \alpha_j\} )$ gathers all vertex factors, propagator numerators, and external spinors in the amplitude. The exponents $\delta_i$ are the powers of the original propagator denominators. As in our previous example, introducing a $q$ dependence  in the integral  by shifting one of the $p_i$'s will result in a function depending on $q$ through invariants of order $O(\lambda^2)$ in the denominator,
\begin{align}
\tilde{F}(p_1, \dots, & \, p_{i_0}+q, \dots, p_n) =  \int  \prod_{l=1}^L d^4 k_l  \int \prod_{j=1}^M d\alpha_j \, \delta\left(1 - \sum_{j=1}^M \alpha_j\right)\times \nonumber \\
&\times \frac{\mathcal{N}(\{p_i, k_l, \alpha_j\},q)}{\left[\sum_{j=1}^M \alpha_j (l_j^2(\{p_i, k_l\}) - m_j^2) + G(\{p_i\cdot q, k_l \cdot q , \alpha_j \} , q^2) + i\epsilon \right]^{\sum_{j=1}^M\delta_j}} \, ,
\end{align}
with $G(\{p_i\cdot q, k_l \cdot q, \alpha_j\} , q^2) = O(\lambda^2)$. If we have
\begin{align} \label{eq:denominator_almost_vanish}
\sum_{j=1}^M \alpha_j (l_j^2(p_i, k_l) - m_j^2) = O(\lambda^2)\, ,
\end{align}
then a power series expansion in $q$ of the denominator becomes inaccurate. Regions in loop variables space where \eqref{eq:denominator_almost_vanish} holds are close to submanifolds where the denominator of the loop integrand in \eqref{eq:generic_loop_integral} vanishes. The latter can be thought of as ``singular submanifolds''. If the integration contour in \eqref{eq:generic_loop_integral} can be deformed away from a singular submanifold by a deviation larger than $O(\lambda^2)$, then a power series expansion in $q$ is possible. This can be achieved as long as the singular submanifold does not coincide with the endpoint of one of the integration contours or is not pinched between pairs of coallescing singularities in the complex plane. Therefore, a necessary condition for having singular or non analytic terms in $q$ in our loop integrals is the presence of \emph{pinch surfaces}.

To summarize, to extend Low's theorem to soft photon emission in the high energy regime, we need to find all pinch surfaces of the loop integrand for the elastic amplitude. These pinch surfaces may yield singular or non analytic terms in the elastic amplitude which prevent us from performing the expansion in $q$ that is crucial to the classic form of Low's argument. Pinch surfaces are found by solving the Landau equations \cite{Landau:1959fi}. Solutions to these equations can be visualized  as physical processes with classical propagation of particles, following an observation first made by Coleman and Norton \cite{Coleman:1965xm}. Physical propagation of on-shell particles is represented using ``reduced diagrams'' where all off-shell lines are shrunk to a point. In general, loop integration over a neighborhood of a pinch surface will yield non analytic logarithmic dependence on the soft momentum $q$ which must be factorized as described below. Not all pinch surfaces result in singular terms however, and in the majority of cases, integration about a given pinch surface will yield a contribution of order higher than is relevant for the soft photon theorem. In the next section, we will use power counting techniques \cite{Libby:1978qf} to determine the order of magnitude of integrals over regions neighboring pinch surfaces, and also to determine if the resulting term is singular or not.




\section{Analytic structure of the radiative and elastic amplitudes}

Low's theorem is an expansion of the soft photon radiative amplitude in powers of $\lambda$ from $O(\lambda^{-2})$ to $O(\lambda^0)$. As explained in Sec.\  \ref{sec:linear_expansion_failure}, incorporating loop corrections in the high energy regime requires a careful study of the loop integrand's pinch surfaces. This is because the loop integration over a region neighboring a pinch surface produces non analytic logarithms of $q$, and in some cases even singular terms at $q =0$.

Following Akhoury and Sen \cite{Akhoury:1978vq,Sen:1982bt}, finding pinch surfaces using the reduced diagrams of Coleman and Norton is straightforward. Once we have found a pinch surface, we will use power counting techniques to put an upper bound on the loop integral over a region close to that pinch surface. This will allow us to determine whether this pinch surface corresponds to a contribution to the soft photon expansion at $O(\lambda^0)$.

Power counting techniques also allow us to determine if the integral about a pinch surface is singular and in fact, these were first introduced to search for infrared singularities in higher loop integrals. We will begin with a brief review of this technology. More detailed treatments are given  in \cite{Collins:1989gx,Sterman:1995fz, Sterman:1994ce, Collins:2011zzd}.

\subsection{Review of power counting} \label{sec:power_counting}

Arbitrary multiloop Feynman diagrams have infrared singularities associated with various limits in their loop integration momenta  \cite{Sterman:1978bi, Sterman:1978bj}. These singularities come from singular submanifolds where propagator denominators vanish. A necessary condition for a singular submanifold to result in a singularity is that it must be a pinch surface. However, a pinch surface need not yield a divergent integral. To determine whether that is the case or not requires the use of power counting techniques \cite{Libby:1978qf}. Power counting allows us to determine the order of growth of a loop integral over a region close to a pinch surface. This procedure is best explained by studying a concrete example.

Consider again the triangle integral in \eqref{eq:power_counting_example}. We consider the pinch surface arising from the  limit where $k$ becomes collinear with $p_1$. To capture how singular this pinch surface is, we need to change the loop integration variables to ``intrinsic'' and ``normal'' coordinates. Normal coordinates are the variables that vanish as we approach the singular submanifold. Intrinsic coordinates, on the other hand, are variables whose variation moves a point along the submanifold without leaving it. 

The scalings in \eqref{eq:basic_scaling} tell us that as we approach the collinear region by taking the limit $\lambda \rightarrow 0$, two denominators vanish as $O(\lambda^2)$, as is required for a singularity. Further, the collinear region is approached by making $k^-$ and $k_T$ small, which leads us to identify these as the normal variables. The remaining large component $k^+$ of $k$ is the intrinsic component. Changing integration variables to the normal and intrinsic coordinates, \eqref{eq:power_counting_example} becomes
\begin{align} \label{eq:power_counting_example_v2}
I =  \pi \int_{c^+}^{b^+} dk^+ \int_{c^- \lambda^2}^{b^- \lambda^2} d k^- \int_{c_T \lambda^2}^{b_T \lambda^2} d k_T^2 & \,  \frac{1}{2 k^+ k^- - k_T^2 + i\epsilon} \nonumber \\
& \times \frac{1}{ 2 k^+ k^- - k_T^2 + 2 p_1^+ k^- + 2 p_1^- k^+ + i\epsilon  } \nonumber \\
& \times \frac{1}{2 k^+ k^- - k_T^2 + 2 p_2^+ k^- + 2 p_2^- k^+ - 2 k_T \cdot p_{2T}+i\epsilon } \, .
\end{align}
In the bounds of integration, we have introduced the numbers $b^*$ and $c^*$ which are all $O(1)$. The bounds include an appropriate power of $\lambda$ since we are interested in the order of magnitude of the loop integral over a region of integration that borders the collinear pinch surface where $k^-$ and $k_T$ vanish. To obtain an estimate of the order of the loop integral about the collinear region, we then perform the changes of variables
\begin{align}
k^+  &= \kappa^+ \nonumber \\
k^- &= \lambda^2 \kappa^- \nonumber \\
k_T &= \lambda \kappa_T \, .
\end{align}
Eq. \eqref{eq:power_counting_example_v2} becomes
\begin{align}
I = \pi \int_{c^+}^{b^+} d\kappa^+ \int_{c^- }^{b^-} \lambda^2 d \kappa^- & \int_{c_T }^{b_T} \lambda^2 d \kappa_T^2  \,  \frac{1}{\lambda^2 (2 \kappa^+ \kappa^- - \kappa_T^2 + i\epsilon)} \nonumber \\
& \times \frac{1}{ \lambda^2 (2 \kappa^+ \kappa^- - \kappa_T^2 + 2 p_1^+ \kappa^- + 2 p_1^-  \kappa^+ / \lambda^2 + i\epsilon)  } \nonumber \\
& \times \frac{1}{2 \lambda^2 \kappa^+ \kappa^- - \lambda^2 \kappa_T^2 + 2 \lambda^2  p_2^+ \kappa^- + 2 p_2^- \kappa^+ - 2 \lambda \kappa_T \cdot p_{2T} + i\epsilon} \, .
\end{align}
Factoring out the leading powers of $\lambda$ in the numerator and denominators, we are left with an overall scaling of $\lambda^0$ times an integral where all three denominators are $O(1)$ and whose domain of integration is well separated from the pinch surface, since its bounds are all $O(1)$. This is the case even though the integration volume vanishes as a power of $\lambda$. Note that the term $2 p_1^- \kappa^+ / \lambda^2$ in the second line is $O(1)$ since the component $p_1^-$ scales as $\lambda^2 = \frac{m^2}{ 2 p_1^+}$. An overall scaling for the integral of $\lambda^0$ indicates the potential for a logarithmic term. In fact, any scaling as a power of $\lambda$ is valid up to multiplication by a logarithmic function of $\lambda$.

The procedure we have just employed to find the potential for a logarithmic divergence without going through a full calculation of a loop integral can be systematized and applied as above to any higher order multiloop diagram for $M_{el}(\{p_i\})$ whenever none of the $p_i$ are parallel. The key points are the identification of candidate pinch surfaces, the proper definition of normal variables and their scaling, and finally power counting to put bounds on the order of growth of the integral. The final step will tell us that the loop integral scales as some power $\lambda^\gamma$ of the small parameter $\lambda$. This power $\gamma$ is called the infrared degree of divergence of the pinch surface, in analogy with the ultraviolet degree of divergence of renormalization theory. A strictly positive degree of divergence $\gamma > 0$ means that we have a non-singular integral. Conversely, a degree of divergence $\gamma \leq 0$ indicates that we have an infrared divergence in the loop integration when $\lambda \rightarrow 0$ i.e. in the massless limit. More specifically, $\gamma < 0$ tells us that the pinch surface leads to a power divergence while $\gamma = 0$ indicates the presence of a logarithmic divergence. In our analysis of soft photon emission, we are interested in retaining a finite mass for the fermion, but we can still apply power counting techniques, as we have described, to determine the order of magnitude of contributions from pinch surfaces to loop integrals.

\subsection{Power counting analysis of $n$ particle scattering} \label{sec:power_counting_applied}

Armed with the tools we have described, we begin our treatment of the soft photon theorem with a  study of the infrared structure of the elastic amplitude.  Although we will only consider outgoing external fermions and antifermions in the interest of conciseness, extending our analysis to include external scalars is straightforward. To classify pinch surfaces based on the order of magnitude of their contribution as a power of $\lambda$, we introduce a separate set of light cone coordinates for each external particle $p_1, \dots, p_n$. As in the example we studied, the normal variables are the transverse and ``minus'' components for a collinear loop momentum, or all momentum components for a soft loop momentum. These definitions of the scaling of momenta in singular regions result in the power counting rules listed in Table \ref{tab:power_counting}. When analyzing  a reduced diagram that represents a given pinch surface, we use the rules in the table to determine the contribution to the infrared degree of divergence from all components of the reduced diagram -- \textit{i.e.} collinear fermion lines, soft fermion lines, etc. Using the Euler identity, it is possible to obtain a general formula for the degree of divergence of the most general reduced diagram \cite{Sterman:1994ce}. Although we will not review the details of such a treatment, we will outline the main intermediate results for convenience.

\begin{figure}[h!]
\centering
\includegraphics[width = 0.75 \textwidth]{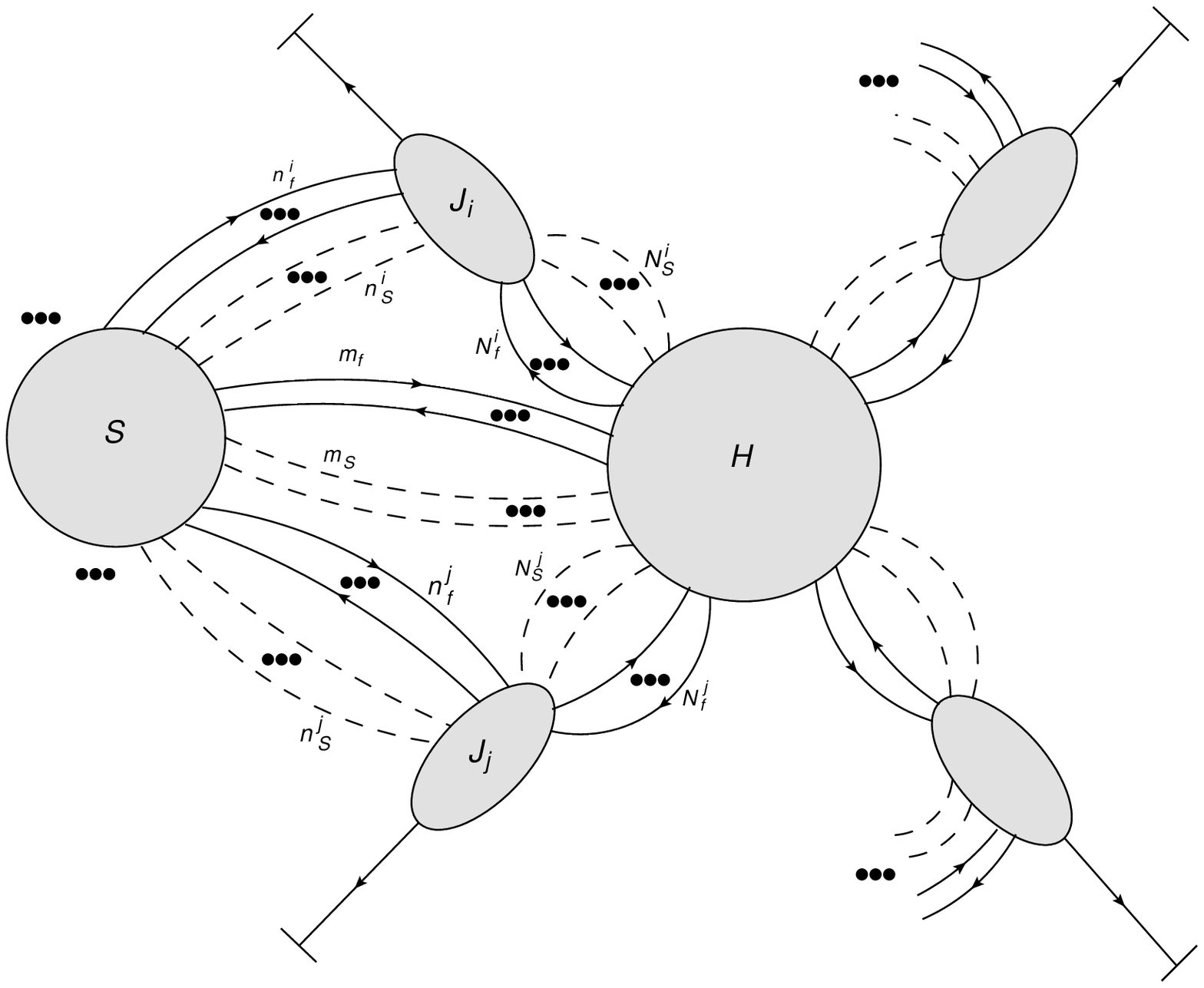}
\caption{The most general reduced diagram incorporating the hard vertex, the soft function, and well separated jets.}
\label{fig:generic_multi_jet}
\end{figure}

As shown in Refs.\ \cite{Akhoury:1978vq, Sen:1982bt}, the application of the Coleman-Norton analysis \cite{Coleman:1965xm} gives the most general reduced diagram for the elastic scattering of $n$ particles, which is shown in Fig.\ \ref{fig:generic_multi_jet}. The hard part labelled $H$ has several jets of collinear particles emerging from it.  In our notation, the jet of lines collinear to the $i^{th}$ external particle is linked to the hard part by $N^i_f$ collinear fermion lines and $N^i_s$ collinear scalar lines. Each jet can also have soft particles emerging from it; for the $i^{th}$ jet, we denote the number of such soft fermions by $n^i_f$ and the number of soft scalars by $n^i_s$. The number $n^i_f+N^i_f$ is odd in the case we study, when the $n^{th}$ external particle is a fermion, and would be even if the external particle were a scalar. The soft fermions and scalars emerge from the $n$ jets and combine at a soft cloud denoted $S$. Finally, there are $m_f$ soft fermions and $m_s$ soft scalars connecting the soft cloud to the hard part. 

\begin{figure}[h!]
\centering
\includegraphics[width = 0.8 \textwidth]{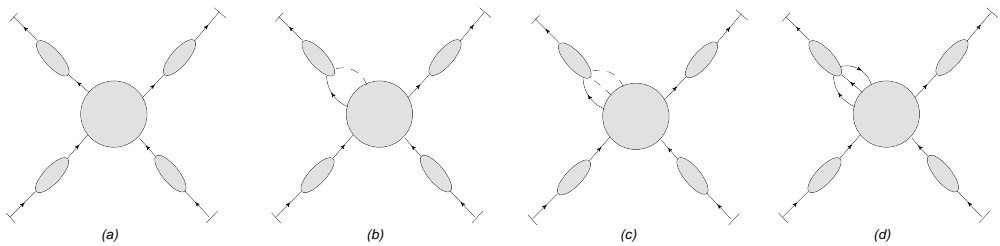}
\caption{Four of the classes of elastic amplitude diagrams that we need to include when attaching a photon for performing Low's analysis. These are the only ones contributing in the massless limit -- see the text for a discussion of this limit. The diagrams in (a) are the leading terms with $\gamma = 0$. Diagrams in (b) have $\gamma =1 $ in the massive case and $\gamma = 2 $ in the massless case. The classes of diagrams in (c) and (d) always have $\gamma = 2$.}
\label{fig:contribution_low_analysis}
\end{figure}

\begin{figure}[h]
\centering
\includegraphics[width = 0.5 \textwidth]{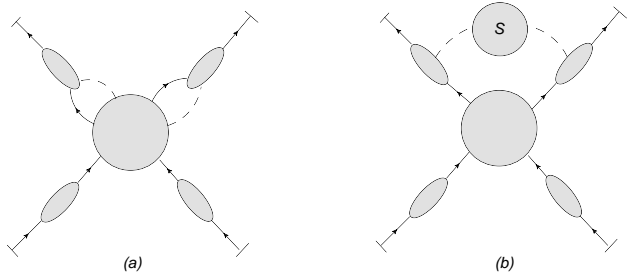}
\caption{In the massive case, we have to include two additional classes of diagrams with $\gamma = 2$. The soft two-point function in (b) includes only soft scalars.}
\label{fig:contribution_massive}
\end{figure}

\begin{figure}[h]
\centering
\includegraphics[width = 0.25 \textwidth]{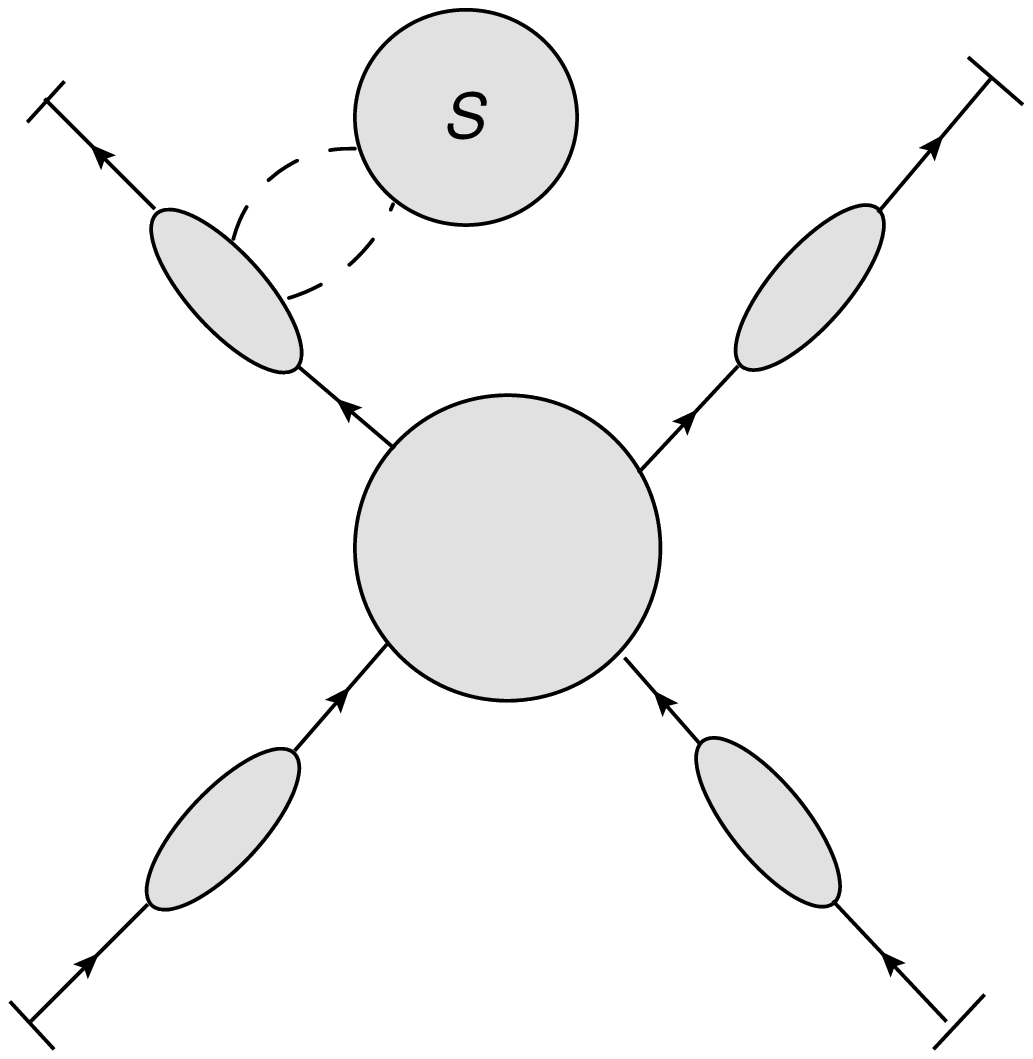}
\caption{This class of diagrams has $\gamma =2$ but will be combined with the leading jet class.}
\label{fig:leading_jet_soft}
\end{figure}

The $i^{th}$ jet's contribution to the degree of divergence of Fig.\ \ref{fig:generic_multi_jet} is denoted by $\gamma_{J_i}$ and the contribution from the soft cloud $S$ will be denoted by $\gamma_S$. Then using the rules from Table \ref{tab:power_counting} and the Euler identity, one can derive the following
\begin{align} \label{eq:power_counting_result}
\gamma_{J_i} &= N^i_f + N^i_s - n^i_f - n^i_s -1 \nonumber \\
\gamma_S &= 4 \sum_i n^i_f + 2 \sum_i n^i_s + I_f  + 4 m_f +  2 m_s   \, ,
\end{align}
where we have introduced the symbol $I_f$ to stand for the number of soft fermion lines internal to the soft cloud $S$. The suppression associated with a Yukawa vertex in Table \ref{tab:power_counting} enters in the derivation of $\gamma_{J_i}$, and follows from the relation $ (\gamma^-)^2 = 0$ and the Dirac equation. The above formulas remain valid whether the $i^{th}$ external particle is a fermion or a scalar. Combining the formulas in \eqref{eq:power_counting_result}, we obtain that the degree of divergence of the most general reduced diagram is
\begin{align} \label{eq:power_counting_final_result}
\gamma &= \sum_i (N^i_f + N^i_s + 3 n^i_f + n^i_s -1) + I_f + 4 m_f + 2 m_S \, .
\end{align}
Using this result, it is possible to identify the diagrams with $\gamma = 0$, $\gamma = 1$, and $\gamma = 2$. These appear in Figs.\  \ref{fig:contribution_low_analysis}, \ref{fig:contribution_massive}, and \ref{fig:leading_jet_soft}. Another consequence of Eq. \eqref{eq:power_counting_final_result} is that there are no diagrams with $\gamma<0$, meaning that the elastic amplitude is at most logarithmically singular in the limit $\lambda \rightarrow 0$. These results were derived long ago by Akhoury in the massless case \cite{Akhoury:1978vq}.


\begin{table}[h!]
\begin{center}
	\begin{tabular}{ |  l | c | c | } 
	\hline
	                                                                 &    Enhancement    &     Suppression \\ \hline
	Collinear fermion line                              &        -2                   &                          \\ \hline
	Collinear scalar line                                &        -2                  &                           \\ \hline
	Soft fermion line                                     &        -1                &                            \\ \hline
	Soft scalar line                                        &         -4                 &                           \\ \hline
	Collinear loop integral                              &                              &           +4           \\ \hline
	Soft loop integral                                     &                              &           +8             \\ \hline
	Yukawa vertex on collinear fermion line  &                              &            +1            \\ \hline
	\end{tabular}
\caption{The power counting rules above define how much each component of a reduced diagram contributes to the degree of divergence of the corresponding pinch surface. These rules are for Yukawa and scalar theories where the fermions are massive and the scalars are massless. In the case of massless fermions, soft fermions yield an enhancement of $-2$ rather than $-1$. }
\label{tab:power_counting}
\end{center}
\end{table}



The regions of integration about the pinch surfaces with $0 \leq \gamma \leq 2$ are precisely the ones we need to consider for extending the soft theorem. To see why, recall first that the soft photon momentum $q$ scales as $(\lambda^2, \lambda^2, \lambda^2)E$. Hence, attaching a soft photon to a  collinear or soft fermion line will not alter the scaling of the pre-existing fermion line. In fact, the net effect as far as our power counting procedure is concerned will be the addition of a supplementary collinear or soft fermion line. Therefore, following the rules in Table \ref{tab:power_counting}, attaching an external soft photon to an elastic amplitude diagram will reduce its degree of divergence by $2$ if the photon is attached to a collinear fermion line, or $1$ if it is attached to a soft fermion line. If we start from diagrams with $0 \leq \gamma \leq 2$, this will leave us with radiative diagrams of order between $O(\lambda^{-2})$ and $O(\lambda^0)$, which is precisely the range of magnitudes relevant to Low's theorem.


The class of diagrams in Fig.\ \ref{fig:contribution_low_analysis}(a) corresponds to the logarithmic leading term of Akhoury. This class consists only of the hard part attached by single fermions to several jets of virtual on-shell lines collinear to the external particles. Since jets attached to the hard part by a single fermion line give the leading term in the elastic amplitude, we will henceforth refer to such jets as ``leading jets''. Adding a soft photon to a diagram with only leading jets generates the leading $O(\lambda^{-2})$ terms in the soft photon theorem. The non-leading jets appearing in Figs. \ref{fig:contribution_low_analysis} (b), (c), and (d) will be referred to as $fs$-jets, $fss$-jets, and $fff$-jets respectively. These labels indicate whether the collinear particles connecting the jets to the hard part are scalars ($s$) or fermions/antifermions ($f$). The two jets connected by a soft two-point function in Fig. \ref{fig:contribution_massive} (b) are named $f\softs$-jets, where the scalar label $s$ is underlined to indicate that it refers to a soft scalar rather than a collinear one.

The diagrams (c) and (d) of Fig.\ \ref{fig:contribution_low_analysis}  have $\gamma = 2$ and inserting a photon into them produces diagrams that are $O(\lambda^0)$ up to multiplication by a logarithmic term in $q$. When the fermions have non zero mass $m$, the diagrams in \ref{fig:contribution_low_analysis}(b) have $\gamma =1$ and inserting a soft photon into them produces a term scaling as $\lambda^{-1}$, thereby yielding a contribution of the same order as $m/q$.

One can take the limit where $m \rightarrow 0$ and retain the definitions of collinear and soft scaling provided another suitable small scale $\lambda$ is identified. We will assume this has been done when discussing the fully ``massless limit'' here and below. In this case, the $fs$-jets scale as $\lambda^2$, unlike the $\lambda$ scaling predicted by power counting in the massive case. This  makes the whole class of diagrams in Fig.\ \ref{fig:contribution_low_analysis}(b) scale as $\lambda^2$.  Hence, the term of intermediate order of magnitude $\lambda^{-1}$ disappears in the massless limit. The $O(\lambda)$ contribution is absent from $fs$-jets in the massless limit because the denominator of the loop integrand is symmetric under simultaneous reflection of all transverse loop momenta while the leading term in the numerator is odd under such a transformation. Consequently, the term that would scale as $\lambda$, and thereby conform to the power counting rules of Table \ref{tab:power_counting}, actually vanishes. This vanishing of the would-be leading term pushes the scaling of $fs$-jets back to the next available order, namely $\lambda^2$. We illustrate how this happens explicitly in Sec.\ $\ref{sec:example_fs}$.

In Fig.\ \ref{fig:contribution_massive}, we show two additional classes of diagrams that only contribute in the massive case. Both of these have $\gamma = 2$. Diagram \ref{fig:contribution_massive}(a) contains two distinct non-leading jets but can be treated similarly to diagrams appearing in Fig.\  \ref{fig:contribution_low_analysis}.  Diagram \ref{fig:contribution_massive}(b) includes a two-point function of soft particles connected to two distinct jets. From the requirement $\gamma \leq 2$ with $\gamma$ given by Eq. \eqref{eq:power_counting_final_result}, we find that this two-point function consists only of soft scalars and therefore no photon is emitted from it. 

The disappearance of contributions from diagrams \ref{fig:contribution_massive}(a) in the massless limit takes place because with massless fermions, $fs$-jets scale as $\lambda^2$ rather than $\lambda$, as we mentioned when discussing Fig.\ \ref{fig:contribution_low_analysis}(b). This makes the diagrams of Fig.\ \ref{fig:contribution_massive}(a) scale as $\lambda^4$ rather than $\lambda^2$. As for contributions from diagrams \ref{fig:contribution_massive}(b), the $f\softs$-jets will scale as $\lambda^{0}$ rather than $\lambda^{-1}$ in the massless case, making the whole diagram scale as $\lambda^{4}$. As in the case of $fs$-jets, the scaling of $f\softs$-jets in the massless limit is due to the leading term in the integrand being odd under simultaneous reflection of all transverse loop momenta.

Finally, Fig.\ \ref{fig:leading_jet_soft} shows a diagram with a soft scalar self energy connecting to a single leading jet. These also have $\gamma = 2$, but since we do not need to consider photon emission from the soft self energy, we will simply include this class of diagrams within the leading jets of Fig.\ \ref{fig:contribution_low_analysis}. 

Now that we have identified all pinch surfaces necessary for extending Low's theorem to high energies, we need to factorize their contributions into jet functions. This will allow us to factor the radiative and elastic amplitude into parts that can or cannot be expanded in a power series in $q$. We turn to this task in Sec. \ref{sec:factorization}.




\section{Factorization of non analytic contributions} \label{sec:factorization}

Our approach to extending Low's analysis will rely on factorization. The effect of the pinches producing jet-like momentum configurations can be captured by universal ``jet functions'' having the same pinch surfaces and singularities as the original Feynman diagrams. The product of jet functions is then matched onto the full amplitude by a ``hard function'' or ``hard part''. The hard part gets its leading contribution from exchanges of hard virtual particles. We assume it can be constructed from an algorithm consisting of nested subtractions similar to the procedure described in \cite{Collins:2011zzd, Collins:1981uk,Erdogan:2014gha}. The jet functions have matrix element definitions, which are closely related to the soft collinear effective theory approach to soft radiation theorems \cite{Larkoski:2014bxa} and the treatment of bound states in \cite{Efremov:1979qk, Lepage:1979zb, Lepage:1980fj}.  In the soft collinear effective theory approach, the role of our hard part is played by the matching coefficients.

To set the stage for our adaptation of Low's analysis, we need to introduce a unifying notation for jet functions. For Akhoury's leading jets, we use the notation $J^f(p_i)$. The superscript $f$ represents the single fermion or antifermion emerging from the hard part and attaching to the collinear lines comprising the jet. The corresponding hard part is simply denoted $H(p_1, \dots, p_n)$. Since leading jets are attached to the hard part by a single fermion line, $J^f(p_i) $ is essentially a reduced on-shell self energy. For example, for an outgoing fermion, $J^f(p_i)$ has the matrix element definition
\begin{align}\label{eq:leading_jet_def}
J^{f} (p_i) &= \langle p_i | \bpsi(0) | 0 \rangle \, .
\end{align}
We now introduce a standard notation appropriate for the factorized amplitude. As is customary when using light cone coordinates, we define the vectors $n_i^\mu$ and $\bar{n}_i^\mu$ by
\begin{align}
\vec{n}_i &= \frac{\vec{p}_i}{\sqrt{2}\, |\vec{p}_i|} = - \vec{\bar{n}}_i\, , \nonumber \\
n_i^2 &= \bar{n}_i^2 =  0 \, , \nonumber \\
n_i \cdot \bar{n}_i &= 1\, .
\end{align}
The vector $n_i$ points in the direction collinear to $p_i$, while the vector $\bar{n}_i$ points in the anti-collinear direction.

The hard part $H$ with which the jets are combined is only sensitive to the collinear components of the external momenta $p_i$. These collinear components are defined by
\begin{align}
\hp_i &= p_i^+ n_i \, .
\end{align}
This collinear vector is the natural argument for the hard part. In the case of a jet loop momentum $k$ collinear to $p_i$, we proceed analogously and define $\hk \equiv k^+ n_i $. In hard parts, the other components of jet loop momenta are set to zero, or in one case expanded about zero (see below).

In the notation just introduced, the leading term in the expansion of the elastic amplitude is
\begin{align}
M_{el}^{leading} &= \left( \prod_{i=1}^n J^f (p_i) \right) \, \otimes  H(\hp_1, \dots, \hp_n) \, .
\end{align}
Each jet function and hard part carries implicit Dirac spinor indices. The tensor product symbol ``$\otimes$'' will stand for a product of Dirac spinors contracted with matching indices in the jet functions and the hard part.

The non leading jet function in Fig.\ \ref{fig:contribution_low_analysis}(b)  is denoted by $J^{fs}(p_i-\hk, \hk)$. The $fs$ superscript indicates that the first momentum in the argument belongs to the collinear fermion connecting the hard part to the jet and the second momentum to the collinear scalar. The corresponding hard part is then denoted $H_i^{fs}(\hp_1,\dots; \hp_i - \hk , \hk; \dots, \hp_n)$. The subscript $i$ and superscript $fs$ indicate that the $i^{th}$ outgoing momentum $p_i$ is split between the collinear fermion momentum $p_i - k$ and the collinear scalar momentum $k$ that are shown between the semicolons for clarity. This $fs$-jet function has the matrix element definition
\begin{align}\label{eq:operator_definition_fs}
J^{fs}(p - \hk, \hk)  &= \int_{-\infty}^{\infty} d\xi \, e^{-i \hk\cdot (\xi \bar{n}_i)} \langle p | \phi(\xi \bar{n}_i) \bpsi(0) | 0 \rangle \, ,
\end{align}
for an outgoing fermion, and the subleading amplitude formed from this jet and hard part has the expression
\begin{align} \label{eq:amplitude_fs_jet}
M^{fs} &= \sum_{i=1}^{n} \left( \prod_{j\neq i} J^f (p_i) \right)  \int_0^{p_i^+} dk^+ \,  J^{fs}(p_i - \hk , \hk)\, \otimes  H_i^{fs}(\hp_1,\dots;\hp_i - \hk , \hk; \dots, \hp_n) \, .
\end{align}
In \eqref{eq:operator_definition_fs}, the argument of the field $\phi$ is $\xi \bar{n}_i$ because we integrate over all non collinear components of the loop momentum $k$. Therefore, $\xi$ is the ``$-$'' component of the original position space argument of the field $\phi$, conjugate to the collinear component $k^+$.

Since $fs$-jets scale as $O(\lambda)$, it is possible to obtain a contribution of order $O(\lambda^2)$ by expanding the hard part $H^{fs}$ to first order in the transverse loop momentum $k_T$ before it is integrated over. This results in contributions captured by a derivative operator and a separate corresponding hard part. The derivative $fs$-jet is denoted $J^{f\partial s}$ and given by the operator definition
\begin{align}
J^{f\partial s} &= \int_{-\infty}^{\infty} d\xi \, e^{-i \hk \cdot (\xi \bar{n}_i)} \langle p | (\partial_T \phi )(\xi \bar{n}_i) \bpsi (0) | 0 \rangle \, .
\end{align}
The transverse index in the derivative is suppressed.  It is contracted with a corresponding index in the matching hard part, which will be denoted by $H^{f\partial s}$. Analogously to \eqref{eq:amplitude_fs_jet}, we have
\begin{align}
M^{f\partial s} &= \sum_{i=1}^{n} \left( \prod_{j\neq i} J^f (p_i) \right)  \int_0^{p_i^+} dk^+ \,  J^{f\partial s}(p_i - \hk , \hk)\, \otimes  H_i^{f\partial s}(\hp_1,\dots;\hp_i - \hk , \hk; \dots, \hp_n) \, ,
\end{align}
for the amplitude with an $f\partial s$-jet, where the symbol $\otimes$ now includes a sum over the transverse index of the derivative, just as for the implicit Dirac indices.

For Fig.\ \ref{fig:contribution_low_analysis}(c), the symbol $J^{fss}(p_i - \hk_1 - \hk_2, \hk_1, \hk_2)$ represents the $fss$-jet where two scalars and a fermion merge at the hard part from a bundle of collinear lines. The momenta $k_1$ and $k_2$ are the momenta of the collinear scalars and $p_i - k_1 - k_2$ is the momentum of the collinear fermion. This jet has the matrix element definition
\begin{align}
J^{fss} (p - \hk_1 - \hk_2, \hk_1, \hk_2) &= \int_{-\infty}^{\infty} d\xi_1 \, e^{-i \hk_1 \cdot (\xi_1 \bar{n}_i)} \int_{-\infty}^{\infty} d\xi_2 \, e^{-i \hk_2 \cdot (\xi_2 \bar{n}_i)} \langle p | \phi(\xi_1 \bar{n}_i) \phi(\xi_2 \bar{n}_i) \bpsi(0) | 0 \rangle \, .
\end{align}
Following the same logic as in the previous case, the corresponding hard part is $H_i^{fss}(\hp_1, \dots; \hp_i - \hk_1 - \hk_2, \hk_1, \hk_2; \dots, \hp_n)$ and the expression for the amplitude with a single $fss$-jet is
\begin{align}
M^{fss} &= \sum_{i=1}^n \left( \prod_{j\neq i} J^f(p_j) \right)  \int_0^{p_i^+} dk_1^+\int_0^{p_i^+} dk_2^+ \, \theta(p_i^+ - k_1^+ - k_2^+) \nonumber \\
&\quad \quad \times J^{fss}(p_i-\hk_1-\hk_2, \hk_1, \hk_2) \otimes  H^{fss}_i(\hp_1, \dots; \hp_i-\hk_1 - \hk_2, \hk_1, \hk_2; \dots,  \hp_n) \, .
\end{align}

Finally, the $fff$-jet function in Fig.\ \ref{fig:contribution_low_analysis}(d) is denoted $J^{fff}(p_i - \hk_1 - \hk_2, \hk_1, \hk_2)$ and its corresponding hard part is $H_i^{fff}(\hp_1, \dots ; \hp_i - \hk_1 - \hk_2, \hk_1, \hk_2; \dots, \hp_n)$. Similarly to the above, this jet has matrix element definition
\begin{align}
J^{fff}(p - \hk_1 - \hk_2, \hk_1, \hk_2)  &= \int_{-\infty}^{\infty} d\xi_1 \, e^{-i \hk_1 \cdot (\xi_1 \bar{n}_i)} \int_{-\infty}^{\infty} d\xi_2 \, e^{-i \hk_2 \cdot (\xi_2 \bar{n}_i)} \langle p | \bpsi(\xi_1 \bar{n}_i) \psi(\xi_2 \bar{n}_i) \bpsi(0) | 0 \rangle \, ,
\end{align}
and the amplitude involving a single $fff$-jet is
\begin{align}
M^{fff} &= \sum_{i=1}^n \left( \prod_{j\neq i} J^f(p_j) \right)   \int_0^{p_i^+} dk_1^+\int_0^{p_i^+} dk_2^+ \, \theta(p_i^+ - k_1^+ - k_2^+)  \nonumber \\
& \quad \quad \times  J^{fff}(p_i-\hk_1-\hk_2, \hk_1, \hk_2) \otimes  H^{fff}_i(\hp_1, \dots; \hp_i-\hk_1 - \hk_2, \hk_1, \hk_2; \dots, \hp_n)\, .
\end{align}
Here, the symbol $\otimes$ includes the contraction of three implicit Dirac indices.

It should be clear at this point that we could generalize our notation to jets with an arbitrary number of particles merging into a jet of collinear lines. It will be convenient to use the same notation for jets of collinear lines in reduced diagrams and the jet functions themselves. Just like the jets themselves, $J^{fs}$, $J^{fss}$, and $J^{fff}$ will also be referred to as $fs$-jets, $fss$-jets, and $fff$-jets respectively. We will also sometimes include integration over the collinear component of loop momenta in the tensor product symbol ``$\otimes$'' and omit the full momentum arguments of jet functions and hard parts when they are clear from the context.

It is interesting to remark that the power-suppressed contributions we have  identified are closely related to exclusive amplitudes for bound states \cite{Efremov:1979qk, Lepage:1979zb,Lepage:1980fj} and next-to-leading power inclusive cross sections for pair production \cite{Kang:2011mg,Kang:2014tta}.

With our notation set up, we can also write down the contribution to the elastic amplitude involving two $fs$-jets,
\begin{align}
M^{fsfs} &= \sum_{1\leq i<j \leq n} \left(\prod_{l\neq i, j} J^f(p_l) \right) \int_0^{p_i^+}dk_1^+  \int_0^{p_j^+}   dk_2^+ \,  J^{fs}(p_i - \hk_1, \hk_1)\nonumber \\
& \quad \quad  \times  J^{fs}(p_j - \hk_2, \hk_2) \otimes H_{ij}^{fsfs}(\hp_1,\dots; \hp_i - \hk_1, \hk_1; \dots ; \hp_j - \hk_2, \hk_2 ; \dots, \hp_n) \, .
\end{align}

As to the $f\softs$-jets from diagrams in Fig.\ \ref{fig:contribution_massive}(b), we use the label $J^{f\softs}(p_i+k,k)$ with the understanding that the momentum argument corresponding to the label $\softs$ is soft rather than collinear. Accordingly, the contribution to the elastic amplitude from diagrams where two $f\softs$-jets are connected by a single soft scalar two-point function $S(k)$ is
\begin{align} \label{eq:f_softs_amplitude}
& M^{f\softs f \softs} = \nonumber \\
&\quad  \sum_{1\leq i<j \leq n} \left(\prod_{l\neq i, j} J^f(p_l) \right)  \int d^4 k \, S(k) J^{f\softs}(p_i + k, k) \, J^{f\softs}(p_j -k, -k)  \otimes H_{ij}^{f\softs f\softs} (\hp_1, \dots, \hp_n) \, .
\end{align}
The hard part has no scalars emerging from it in this case. As we mentioned previously,  Eq. \eqref{eq:power_counting_final_result} implies that $S(k)$ only contains soft internal scalars and cannot radiate any photon at $O(\lambda^0)$. The matrix element definition of the $f\softs$-jet in Eq.\ \eqref{eq:f_softs_amplitude} is
\begin{align}
J^{f\softs}(p + k , k ) &= \int d^4 y \, e^{i k \cdot y} \langle p | \, \frac{\delta S_I}{\delta \phi(y)} \,  \bpsi(0) | 0 \rangle \, ,
\end{align}
where $S_I = \int d^4 x \, \mathcal{L}_I$.  In Yukawa theory, $\mathcal{L}_I(x) = g\, \phi(x) \bpsi(x) \psi(x) + \frac{g^\prime}{4!} \,  \phi^4(x)$, where $g$ is the Yukawa coupling and $g^\prime$ is the four-scalar coupling. No component of the soft momentum $k$ is integrated out. Also, the loop momentum $k$ bears no hat in the above because it is a soft rather than a collinear momentum.

Combining the above definitions allows us to write the fully factorized elastic amplitude,
\begin{align} \label{eq:full_non_radiative}
M_{el} = & \left(\prod_{i=1}^n \tensor{J}{_i^f}\right)\otimes  H  \nonumber \\
&+ \sum_{i=1}^n \left(\prod_{j \neq i} \tensor{J}{_j^f}\right) \tensor{J}{_i^f^s} \otimes \tensor{H}{_i^f^s}+ \sum_{i=1}^n \left(\prod_{j \neq i} \tensor{J}{_j^f}\right) \tensor{J}{_i^f^\partial^s} \otimes \tensor{H}{_i^f^\partial^s} \nonumber \\
&+ \sum_{i=1}^n \left(\prod_{j \neq i} \tensor{J}{_j^f}\right) \tensor{J}{_i^f^s^s} \otimes \tensor{H}{_i^f^s^s} + \sum_{i=1}^n \left(\prod_{j \neq i} \tensor{J}{_j^f}\right) \tensor{J}{_i^f^f^f} \otimes \tensor{H}{_i^f^f^f} \nonumber \\
&+ \sum_{1\leq i < j \leq n} \left( \prod_{l\neq i,j} \tensor{J}{_l^f} \right) \tensor{J}{_i^f^s} \tensor{J}{_j^f^s} \otimes H_{ij}^{fsfs} + \sum_{1\leq i < j \leq n} \left( \prod_{l\neq i,j} \tensor{J}{_l^f} \right) \tensor{J}{_i^f^\softs} \tensor{J}{_j^f^\softs} \, S \otimes  H_{ij}^{f\softs f\softs}  \nonumber \\
& + O (\lambda^3) \, .
\end{align}
When we derive the small $q$ expansion of the radiative amplitude, we will have to consider emission from each factor in each term: the leading jets, the non leading jets, and the hard part. These emission amplitudes are described by the radiative jet functions $ J_{i,\mu}^f (p_l + q , q )$, $ J_{i,\mu}^{fs}(p_i + q - \hk , \hk,q )$, etc.,  and the radiative hard parts $H_\mu(\hp_1, \dots, \hp_n, q)$, $ H_{i,\mu}^{fs}(\hp_1,...,\hp_i-\hk,\hk,...,\hp_n,q)$, etc.  The notation is the same as for non radiative jet functions and hard parts, with the photon index $\mu$ coupling to the polarization of the emitted photon. The radiative jet functions are derived from the matrix element definitions above by inserting an electromagnetic current operator \cite{DelDuca:1990gz}. For example, the elastic jet function in \eqref{eq:leading_jet_def} becomes the radiative leading jet
\begin{align}
J^{f,\mu} (p_i, q) &= \int d^4 x \, e^{i q \cdot x}\langle p_i | j^\mu(x) \bpsi(0) | 0 \rangle \, ,
\end{align}
where $j^{\mu}(x)$ is the electromagnetic current operator and $q$ is the photon momentum.




\section{Examples}

Our discussion of the soft photon theorem above relies on a treatment of the analytic structure of soft photon radiation. In particular, we have used a power counting analysis to determine that non leading jets do contribute to Low's theorem. It is instructive to verify our claims by studying  explicit examples. To this end, we study the lowest order $fs$-jet, and a diagram with two one-loop $f\softs$-jets. We will consider a massless pseudoscalar coupled to massive fermions in this section.  A pseudoscalar coupling lets us use the Dirac equation to obtain more compact formulas for the jet functions but still obeys our power counting rules shown in Sec.\ \ref{sec:power_counting_applied}.

\subsection{Lowest order $fs$-jet}\label{sec:example_fs}

\begin{figure}
\centering
\includegraphics[width = 0.25 \textwidth]{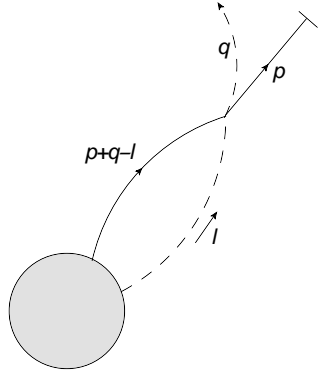}
\caption{Lowest order $fs$-jet after the application of the Ward identity. The photon momentum exits the jet at a composite scalar-photon-fermion vertex as in the Ward identity shown in Fig. \ref{fig:compton_ward_identity}.}
\label{fig:lowest_fs}
\end{figure}

Consider first the lowest order $fs$-jet in a pseudoscalar theory, shown in Fig.\ \ref{fig:lowest_fs}. In the classic form of Low's argument, we would encounter this jet when deducing the internal emission amplitude from the external amplitude as in Eq. \eqref{eq:low_insight_explicit} -- see Fig. \ref{fig:compton_ward_identity}. In Eq. \eqref{eq:low_insight_explicit}, the elastic amplitude is given a $q$ dependence by shifting one of the external momentum arguments $p$ to $p+q$. We have introduced such a $q$ dependence in our $fs$-jet as well, and in doing so aim at exhibiting the logarithmic dependence on $q$ that we described in Sec. \ref{sec:linear_expansion_failure}.
%
%
%
%
%
%
%
%
%
%

The loop integral for the elastic amplitude with a single $fs$-jet takes the form
\begin{align}
M^{fs}(p+q) &= \int_0^1 dx \int \frac{d^d l}{(2\pi)^d} \, \bar{u}(p) \frac{ (-i g \mu^\epsilon \gamma_5 ) i(\pslash + \qslash - \slashed{l} + m ) }{((p+q-l)^2 - m^2 + i\epsilon) } H^{fs}(x)\, \frac{i}{ l^2 +  i\epsilon }\, p^+ \delta(l^+ - x p^+) \, .
\end{align}
We work in dimension $d = 4 - 2 \epsilon$. Our coupling constant for Yukawa theory is $g \mu^\epsilon$, where we introduce the mass scale $\mu$ to retain a dimensionless coupling.

The hard part as we have presented it in Sec. \ref{sec:factorization} depends on the collinear components of the fermion and pseudoscalar momenta: $H^{fs} = H^{fs}( \hp - \hat{l}, \hat{l})$. We can choose our coordinates such that the collinear parts of $p$ and $l$  coincide with their $+$ components. Then, through the delta function $\delta(x -  l^+/p^+) = p^+ \delta(l^+ - x p^+)$, we introduce the new integration variable $x$. This is the fraction of the collinear component of the fermion momentum $p$ taken by the collinear scalar. Since the hard part is only sensitive to the collinear components of its arguments, $H^{fs}$ is a function of $x$ only. Introducing $x$ allows us to integrate over the whole range of loop momenta $l$ while leaving the $x$ integration undone.

The loop integral for $M^{fs}$ can be evaluated analytically without making any simplifying assumption. However, for our purposes, it is convenient to use the frame chosen above with $p_T = 0$. The result of the integration over loop momenta $l$ after retaining the leading term only is then
\begin{align} \label{eq:elastic_fs_jet}
M^{fs}(p+q) = \frac{-gm \mu ^{-\epsilon}}{(4 \pi)^{2-\epsilon}} \Gamma(\epsilon) \int_0^1 dx \, x\,  \bar{u}(p) \gamma_5 H^{fs}(x) \left(x^2 \frac{m^2}{\mu^2} - 2 x (1-x) \frac{p^+ q^- }{\mu^2}\right)^{-\epsilon} \, ,
\end{align}
which is of order $\lambda$ for $\epsilon = 0$, as predicted by the power counting rules of Sec. \ref{sec:power_counting_applied}. The presence of the factor $\Gamma(\epsilon)$ indicates that there is an ultraviolet divergence coming from the loop integral in the definition of the jet function. Therefore, the jet function must be renormalized and thereby becomes a scale dependent quantity. However, the hard part must also be renormalized so that the factorized amplitude matches the original amplitude. This induces evolution equations for the jet functions and the hard parts, as for the treatment of bound states \cite{Efremov:1979qk, Lepage:1979zb, Lepage:1980fj} and in soft collinear effective theory \cite{Larkoski:2014bxa}.

If we apply an on-shell renormalization scheme and subtract the $q=0$ part of the non radiative $fs$-jet in \eqref{eq:elastic_fs_jet}, we obtain at order $\epsilon^0$
\begin{align} \label{eq:elastic_fs_jet_evaluated}
M^{fs}(p+q) = \frac{gm}{(4 \pi)^{2}} \int_0^1 dx \, x\,  \bar{u}(p) \gamma_5 H^{fs}(x) \log\left(1- 2 \left(\frac{1-x}{x}\right) \frac{p^+ q^- }{m^2}\right)\, ,
\end{align}
which exhibits the logarithmic dependence on $q$ that we predicted in Sec. \ref{sec:linear_expansion_failure}. A standard analysis following Low would treat radiative $fs$-jets as part of the internal emission amplitude and deduce their values by expanding \eqref{eq:elastic_fs_jet_evaluated} to linear order in $q$.  This is inaccurate for a photon momentum in the region $q \sim \lambda^2 E$ since we then have $\frac{p^+ q^- }{m^2} \sim 1$, thereby precluding an expansion of $ \log\left( 1 - 2\left(\frac{1-x}{x}\right) \frac{p^+ q^-}{m^2} \right)$ in powers of $q$. We therefore confirm that we are required to include radiative $fs$-jets in the \emph{external} amplitude.

\begin{figure}
\centering
\includegraphics[width = 0.25 \textwidth]{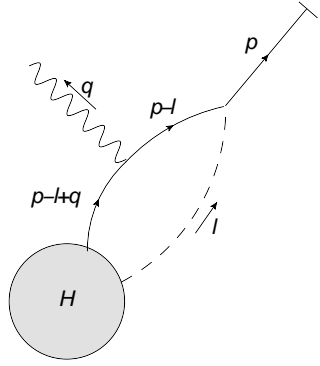}
\caption{Lowest order radiative $fs$-jet.}
\label{fig:lowest_radiative_fs}
\end{figure}

Another prediction of our power counting rules is that adding a photon to the internal collinear fermion line of the non radiative $fs$-jet reduces its degree of divergence by 2. We may verify this expectation by considering the radiative $fs$-jet shown in Fig.\ \ref{fig:lowest_radiative_fs}. 
%
%
%
%
%
%
%
The expression we need to calculate the radiative $fs$-jet function is
\begin{align}
M_{ext}^{fs,\mu} = & \int_0^1 dx \int \frac{d^dl}{(2\pi)^d} \, \bar{u}(p) (-i g \mu^\epsilon \gamma_5) \frac{ i(\pslash- \slashed{l} + m)}{(p-l)^2 - m^2 + i\epsilon} (-i e \mu^\epsilon \gamma^\mu) \times  \nonumber \\
 & \times \frac{i(\pslash - \slashed{l}+ \qslash + m)}{(p-l+q)^2 - m^2 + i\epsilon} H^{fs}(x) \frac{i}{l^2+ i\epsilon} p^+ \delta(l^+ - x p^+) \, .
\end{align}
This can also be evaluated analytically without making any simplifying assumption. However, we do not need to go into that amount of detail to confirm the results of  power counting and will, as before, keep the fermion moving in the $+$ direction. Retaining only the leading term yields the expression
\begin{align}
M_{ext}^{fs,\mu} =  -\frac{eg m }{32 \pi^2 q^-} \int_0^1 dx \, x \, \bar{u}(p) \gamma_5 \gamma^\mu \gamma^- H^{fs}(x) \,  \log\left( 1 - 2\left(\frac{1-x}{x} \right) \frac{p^+ q^-}{m^2} \right) \,.
\end{align}
The order of the radiative $fs$-jet is $\lambda^{-1}$ as predicted by power counting. This is qualitatively new since  there are no terms of order $O(\lambda^{-1})$ in the classic form of Low's theorem. The factor of magnitude $O(\lambda^{-1})$  appears as $m/q^-$ in our example.

We conclude our study of the lowest order $fs$-jet by confirming that in the fully massless case, $m=0$, the non-radiative $fs$-jet is pushed back to $O(\lambda^2)$. To see this, we consider the lowest order $fs$-jet loop integral after the massless condition has been implemented,
\begin{align}
M^{fs}(p) =  \int_0^1 dx \int \frac{dl^+ dl^- d^{d-2}l_T}{(2\pi)^d} & \, \frac{ \bar{u}(p)( -i g \mu^\epsilon \gamma_5 )i(- l^+ \gamma^- - l^- \gamma^+ + \gamma_T \cdot l_T) H^{fs}(x)}{l^2 - 2 p\cdot l + i\epsilon} \nonumber \\
& \times \frac{i}{l^2 + i\epsilon}\,  p^+ \delta(l^+ - x p^+) \, .
\end{align}
Since $p_T = 0$, the denominator of the integrand is even in $l_T$ and therefore the transverse term in the numerator can be ignored. Further, from the massless Dirac equation, $\bar{u}(p) \gamma^- = 0$ when $p_T = 0$, which implies that the leading term in the numerator is $O(\lambda^2)$ since $\gamma^+ l^- = O(\lambda^2)$. It is then a simple matter to finish estimating the magnitude of the integral and conclude that it is $O(\lambda^2)$. This result can be extended to arbitrary order $fs$-jets by proving that any term in the integrand scaling as an odd power of $\lambda$ must be odd under simultaneously reversing the sign of every transverse momentum integration variable.

\subsection{One-loop jets with soft line} \label{sec:example_soft}

\begin{figure}
\centering
\includegraphics[width = 0.75 \textwidth]{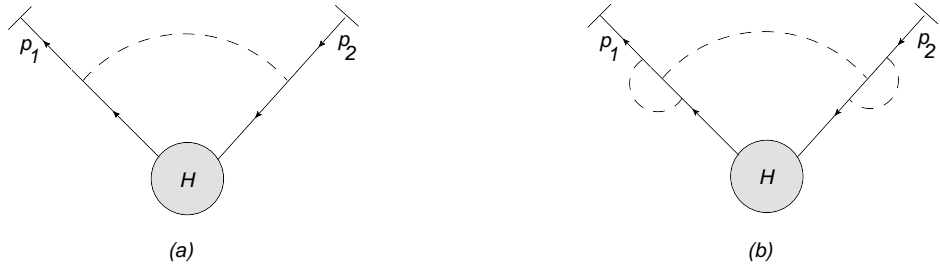}
\caption{Diagram (a) shows the leading order (in $g$) case where a soft pseudoscalar connects two leading jets. This diagram is too suppressed to affect our extension of Low's theorem but diagram (b) is not.}
\label{fig:example_soft}
\end{figure}

The next example we study belongs to the class of diagrams where two $f\softs$- jets are connected by a two-point function of soft scalars such as in Fig.\ \ref{fig:contribution_massive}(b).  In pseudoscalar theories, the leading order diagram in the Yukawa coupling $g$ with two $f\softs$-jets shown in Fig.\ \ref{fig:example_soft}(a) has degree of divergence  $\gamma > 2$ and hence does not contribute to Low's theorem when a soft photon is attached to it. However, this does not extend to higher loop diagrams, as we demonstrate through this section's example.


\begin{figure}
\centering
\includegraphics[width = 0.25 \textwidth]{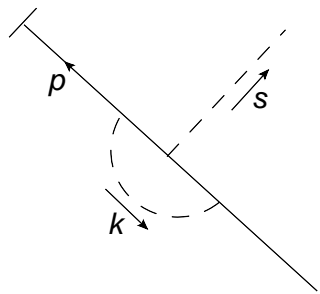}
\caption{The momenta assignments for the one-loop $f\softs$-jet that we are considering.}
\label{fig:higher_order_fsbar}
\end{figure}


Consider then the diagram shown in Fig.\ \ref{fig:example_soft}(b). This diagram has a single soft scalar connecting two one-loop $f\softs$-jets. We consider each $f\softs$-jet individually -- see Fig.\ \ref{fig:higher_order_fsbar}. The expression for a single $f\softs$-jet is
\begin{align}
J^{f\softs}& (p,s) = \nonumber \\
& \int \frac{d^d k}{(2\pi)^d}  \, \frac{\bar{u}(p)( - i g \mu^\epsilon \gamma_5) i(\pslash + \kslash + m)( - i g \mu^\epsilon \gamma_5) i( \pslash + \kslash + \sslash + m ) ( - i g \mu^\epsilon \gamma_5) i(\pslash+\sslash + m)}{((p+k )^2 - m^2)((p+k+s)^2 - m^2)((p+s)^2-m^2)} \frac{i}{k^2}  \, .
\end{align}
The momentum $s$ is the connecting soft pseudoscalar momentum. The loop momentum $k$ is the collinear pseudoscalar momentum. The leading term of this jet function in four dimensions is
\begin{align}
J^{f\softs} (p,s) &=  \frac{g^3}{32 \pi^2}\, \frac{m}{p\cdot s} \, \bar{u}(p) \gamma_5 \left[ \mathrm{Li}_2 (1+2a) + \frac{2a}{1+2a} \log(-2a) - \frac{\pi^2}{6} \right] \bigg|_{a = \frac{p\cdot s }{m^2}} \, ,
\end{align}
which is $O(\lambda^{-1})$. Returning to Fig.\ \ref{fig:example_soft}(b), the expression for the full diagram is
\begin{align}\label{eq:full_fsofts_jet}
M^{f\softs f \softs}(p_1, p_2) &= \int d^4 s \, \frac{i}{s^2 + i \epsilon} J^{f\softs}( \hp_1, s) \, J^{ f\softs}(-\hp_2, -s) \otimes  H^{f \softs f \softs}(\hp_1 + s, -\hp_2 - s)\, .
\end{align}
Bearing in mind that when calculating the degree of divergence, a soft loop integration measure contributes a suppression of $+8$ while a soft pseudoscalar propagator supplies an enhancement of $-4$,  we find that the entire amplitude in \eqref{eq:full_fsofts_jet} has degree of divergence  $\gamma = 8 \times 1 - 1 \times 2 - 4 \times 1 = 2$. This result is the same as if we had proceeded by applying power counting rules to the diagram directly without going through an explicit calculation of the $f\softs$-jet function. Consequently, attaching a soft photon to diagram \ref{fig:example_soft}(b) will yield a correction to the soft theorem of order $O(\lambda^0)$ possibly multiplied by non analytic polylogarithms. Finally, as in the case of $fs$-jets, the leading term of $f\softs$-jets vanishes in the massless fermion limit because its integrand is odd under reflection of all transverse loop momenta.




\section{Adapting Low's argument to the factorized amplitude} \label{sec:low_argument}

In this section, we will first show a preliminary version of our extension of Low's theorem and see that to order $O(\lambda^0)$, we need only consider the hard part corresponding to diagrams with leading jets. Photon emission from hard parts connecting to non leading jets will also be derived for completeness. This treatment adapts Low's argument to our factorized elastic amplitude \eqref{eq:full_non_radiative}.

\subsection{Preliminary form of Low's theorem} \label{sec:low_preliminary}

\begin{figure}
\centering
\includegraphics[width = 0.8 \textwidth]{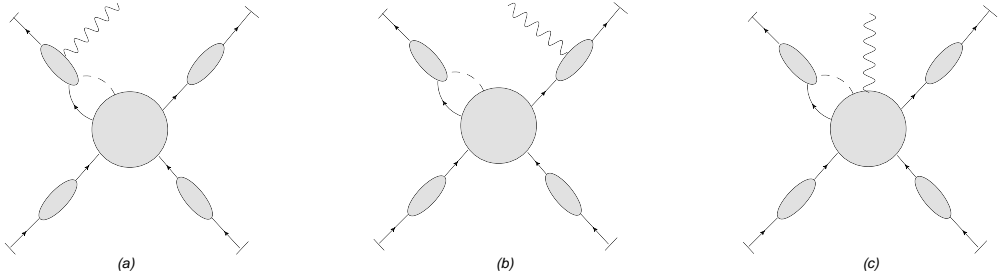}
\caption{In applying Low's argument, the radiative amplitude diagrams are split between those with external photon emission as in (a) and (b), and those with internal photon emission as in (c).}
\label{fig:low_argument}
\end{figure}

To set the stage for the appearance of our preliminary version of Low's theorem,  we first represent the full radiative amplitude in the generic form
\begin{align} \label{eq:generic_radiative}
M_\mu &= \sum_{i=1}^n \left(\prod_{j\neq i} J_j^f \right) J_{i,\mu}^f \otimes H + \left(\prod_{i=1}^n J_i^f \right) \otimes H_\mu \nonumber \\ 
& \quad +\sum_{\theta \in \Theta_1} \sum_{i=1}^n \left[   \left(\prod_{j\neq i} J_j^f \right) J_{i,\mu}^\theta \otimes H_i^\theta + \sum_{l \neq i } \left( \prod_{j \neq i, l} J_j^f\right) J_{l,\mu}^f J_i^\theta \otimes H_i^\theta \right] \nonumber \\
&\quad + \sum_{\theta \in \Theta_2} \sum_{i \neq j} \left[ \left(\prod_{l \neq i,j} J_l^f \right) J_{i,\mu}^\theta J_j^\theta \,  S^\theta \otimes  H_{ij}^{\theta\theta}  + \frac{1}{2} \sum_{h \neq i,j} \left(\prod_{l \neq i,j,h} J_l^f\right) J_i^\theta J_j^\theta J_{h,\mu}^f \,  S^\theta \otimes  H_{ij}^{\theta\theta}   \right]\nonumber \\
&\quad + O(\lambda)\, ,
\end{align}
where $S^\theta =1$ if $\theta = fs$ and $S^\theta = S$ if $\theta = f\softs$. The symbols $\Theta_1$ and $\Theta_2$ stand for the sets of labels $\{fs, f\partial s,  fss, fff \}$ and $\{ fs, f\softs \}$ respectively. Comparing with Eq.\ \eqref{eq:full_non_radiative}, one sees that each term corresponds to a factorized form describing photon emission from a leading jet, a non leading jet, or a hard part -- as illustrated in Fig.\ \ref{fig:low_argument}. Notice however, that we have omitted the following radiative contributions,
\begin{align}\label{eq:higher_order_internal}
E_\mu \equiv \sum_{\theta \in \Theta_1} \sum_{i=1}^n \left(\prod_{j \neq i} J_j^f \right) J_i^\theta \otimes H_{i,\mu}^\theta  + \frac{1}{2}\sum_{\theta \in \Theta_2} \sum_{i \neq j}\left(\prod_{l \neq i,j} J_l^f \right) J_i^\theta J_j^\theta \,  S^\theta \otimes H_{ij,\mu}^{\theta\theta} \, .
\end{align}
These correspond to attaching a photon to the hard parts of the diagrams from Figs. \ref{fig:contribution_low_analysis} (b), (c), (d), and \ref{fig:contribution_massive}, all of which have $\gamma > 0$. Attaching a photon to a hard line does not modify the degree of divergence of the diagram. Therefore, the radiative contributions in \eqref{eq:higher_order_internal} are of order higher than $O(\lambda^0)$ and do not need to be included in our extension of Low's theorem. 


The only radiative hard part contributing at our order in $\lambda$ is the one corresponding to the diagram with leading jets only. We adapt Low's analysis to this diagram by introducing new notation for the external and internal amplitudes,
\begin{align}
M_{ldg}^{ext,\mu} &= \left( \prod_{j\neq i} J_j^f  \right) J_{i}^{f,\mu} \otimes  H \nonumber \, , \\
M_{ldg}^{int,\mu} &= \left( \prod_{i=1}^n J_i^f  \right) \otimes H^\mu  \, .
\end{align}
As before, these are related by the Ward identity,
\begin{align} \label{eq:ward_for_factorized}
q_\mu M_{ldg}^{ext,\mu} + q_\mu M_{ldg}^{int,\mu} = 0 \, ,
\end{align}
which we can use to deduce the radiative hard part $H^\mu$. Indeed, once the jets have been factorized as in Eqs.\ \eqref{eq:generic_radiative} and \eqref{eq:higher_order_internal},  the non radiative and radiative hard parts can be reliably expanded in powers of $q$ even in the regime $q \sim \lambda^2 E$. We emphasize that it is possible to expand the hard parts because the  jet functions are engineered to contain all the leading and next to leading pinch surfaces  of the original Feynman diagrams. All the infrared singularity structure of the radiative amplitude is contained within the jet functions. As noted above, the hard parts correspond to the matching coefficients of soft collinear effective theory \cite{Larkoski:2014bxa} and are assumed to be constructible by nested subtractions similar to \cite{Collins:2011zzd,Collins:1981uk, Erdogan:2014gha}. Hence, hard parts get their leading contributions from off-shell lines and are dominated by hard momenta in the factorized form. Adding a $q$ dependence to hard lines will produce only subleading behavior. This subleading behavior can be represented by higher order terms in an expansion of the hard line propagators in $q$.

It is important to bear in mind that the Ward identity holds not only diagram by diagram, but also at fixed loop momenta, and therefore pinch surface by pinch surface. By this we mean that given a single reduced diagram contributing to an elastic scattering amplitude, the Ward identity will apply to this diagram by itself if we sum over all points of photon insertion \cite{Sterman:1994ce}. The only exception arises in fermion loops where a shift of the loop momentum by the photon momentum $q$ is required.   Therefore, in \eqref{eq:ward_for_factorized}, it is not necessary to include the contributions from all reduced diagrams at once. Rather, we can focus on each class of diagrams individually.

In the case of non leading jets, say an $fs$-jet for definiteness, the internal radiative amplitude $M_{fs}^{int,\mu}$ is of order $O(\lambda)$. The external amplitude $M_{fs}^{ext,\mu}$, on the other hand, is of order $O(\lambda^{-1})$, and the Ward identity has the form
\begin{align}
q^\mu M_{fs}^{ext,\mu} = - q^\mu M_{fs}^{int,\mu} \, .
\end{align}
A naive estimate tells us that the left hand side is $O(\lambda)$ while the right hand side is $O(\lambda^3)$. Therefore, leading terms in the external amplitudes must cancel each other in the Ward identity. In Sec. \ref{sec:higher_order_internal_hard}, we will see that this necessary cancellation also follows from charge conservation -- see the discussion below \eqref{eq:equation_solve_hard}.

\begin{figure}
\centering
\includegraphics[width = 0.8 \textwidth]{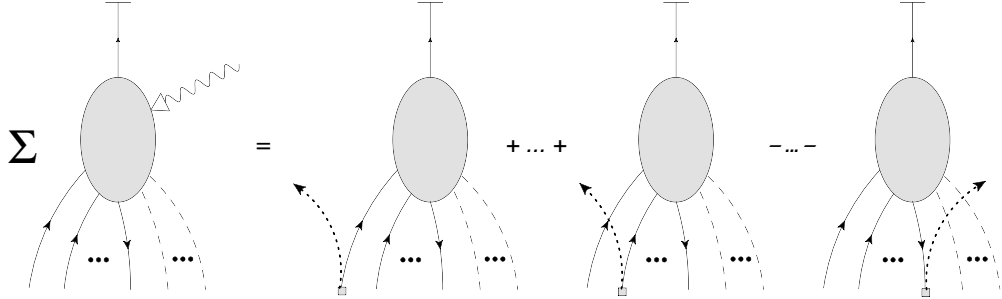}
\caption{The general diagrammatic form of the Ward identity for jet functions. Each fermion line that does not form a closed loop has a term corresponding to the photon exiting the jet at the end of the line and a term for the photon exiting at the beginning of the line. The exception to this rule is the through going fermion line that becomes the outgoing fermion -- this line only has a term with the soft photon exiting the diagram from the beginning and not at the on-shell external line. The terms with the photon exiting at the beginning of a fermion line appear in the identity with a relative $+$ sign whereas those where the photon exits at the end of fermion line carry a $-$ sign.}
\label{fig:jet_ward_identity}
\end{figure}

If we were to try and calculate the internal radiative amplitude $ M_{int}^\mu$ directly, we would need to delve into the construction of the hard part at each order and find all ways of inserting a soft photon, thereby making the calculation different for each process. The alternative, following Low,  is to calculate the contraction of the soft momentum $q$ with the universal external radiative amplitudes and then use the Ward identity to extract  $M_{int}^{\mu}$. However, to detemine $q_\mu M_{ext}^{\mu}$ , we still need to express the quantity  $q_\mu  J_i^{f,\mu}$ in a form that also does not depend on the details of the radiative jet function. To this end, we introduce the jet Ward identities. In their most general form, these are expressed diagrammatically as in Fig.\ \ref{fig:jet_ward_identity}.


The jet QED Ward identity for photon emission is derived straightforwardly using diagrammatic or path integral techniques \cite{'tHooft:1971fh, Sterman:1994ce}. The special case we need involves the leading jet,
\begin{align}
q_\mu J^{f,\mu}(p_i + q , q ) &= e_i J^f(p_i) \, .
\end{align}
It is then straightforward to extend Low's argument and obtain
\begin{align} \label{eq:leading_radiative_hard}
H^\mu(\hp_1, \dots, \hp_n, q) &= - \sum_{l=1}^n e_l \frac{\partial}{\partial \hp_{l\mu}}H(\hp_1^\prime, \dots, \hp_n^\prime) + O(\lambda^2) \, ,
\end{align}
for the radiative hard part. The momentum arguments $\hp_i^\prime$ indicate that we have performed the construction described in Sec. \ref{sec:review_low_theorem} to transition to an elastic set of momenta.

With \eqref{eq:leading_radiative_hard}, we can write the preliminary form of our extension of Low's theorem to high energies.
Incorporating the  radiative hard part \eqref{eq:leading_radiative_hard} into the generic radiative amplitude \eqref{eq:generic_radiative}, we find
\begin{align}\label{eq:preliminary_low}
M_\mu &=   - \left( \prod_{i=1}^n J_i^f \right)  \otimes \sum_{l=1}^n e_l \frac{\partial}{\partial \hp_l^\mu} H \bigg|_{P_0} \nonumber \\
&\quad + \sum_{i=1}^n \left(\prod_{j\neq i} J_j^f \right) J_{i,\mu}^f \otimes  H \nonumber \\ 
&\quad  +\sum_{\theta \in \Theta_1} \sum_{i=1}^n \left[   \left(\prod_{j\neq i} J_j^f \right) J_{i,\mu}^\theta \otimes H_i^\theta + \sum_{l \neq i } \left( \prod_{j \neq i, l} J_j^f\right) J_{l,\mu}^f J_i^\theta \otimes H_i^\theta \right] \nonumber \\
&\quad + \sum_{\theta \in \Theta_2} \sum_{i \neq j} \left[ \left(\prod_{l \neq i,j} J_l^f \right) J_{i,\mu}^\theta J_j^\theta \,  S^\theta \otimes  H_{ij}^{\theta\theta}  + \frac{1}{2} \sum_{h \neq i,j} \left(\prod_{l \neq i,j,h} J_l^f\right) J_i^\theta J_j^\theta J_{h,\mu}^f \,  S^\theta \otimes H_{ij}^{\theta\theta}   \right]\nonumber \\
&\quad + O(\lambda)\, .
\end{align}
We use the $P_0$ symbol to denote that after the momentum derivatives have acted on the non radiative hard part, we evaluate the resulting expression at an elastic set of momenta close to the starting radiative configuration, as described in Sec. \ref{sec:review_low_theorem}. The first term gathers all internal emission at $O(\lambda^0)$. It is remarkable that even at this level, summing over all insertions of a soft photon into the hard part results in the action of a differential operator acting on its external momenta. On the other hand,  as has been mentioned in Sec. \ref{sec:review_low_theorem}, this formula does not yet fully clarify the structure of the radiative jet functions. In contrast to the internal emission, these are universal, and for soft $q$, their structure can be probed using Grammer and Yennie's $KG$ decomposition. We will turn to this task in Sec. \ref{sec:kg_decomposition}.

\begin{figure}
\centering
\includegraphics[width = 0.25 \textwidth]{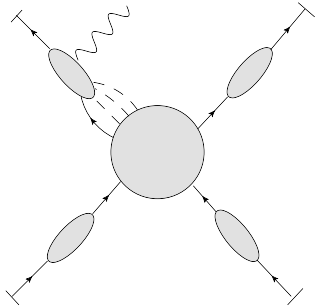}
\caption{An example of a higher order jet required for a consistent treatment of the radiative amplitude $E_\mu$ in \eqref{eq:higher_order_internal}.}
\label{fig:higher_order_external}
\end{figure}

For completeness, we show next how we can adapt Low's insight to analyze the nonleading radiative amplitude $E_\mu$ in \eqref{eq:higher_order_internal}. This amplitude could in fact be included in a higher power treatment of Low's theorem, which would  include higher order jets such as the one shown in Fig.\  \ref{fig:higher_order_external}, and also depend on the higher order subleading terms in the hard part.


\subsection{Photon emission beyond $O(\lambda^0)$} \label{sec:higher_order_internal_hard}

Consider first the class of diagrams with a single $fs$-jet.  For the diagrams in Fig.\ \ref{fig:contribution_low_analysis}(b), the factorized amplitude consists in the $fs$-jet, the leading jets, and the hard part. As shown in Eqs. \eqref{eq:generic_radiative} and \eqref{eq:higher_order_internal}, a soft photon can be emitted from any of those components. The corresponding radiative amplitudes from Eqs.\ \eqref{eq:generic_radiative} and \eqref{eq:higher_order_internal} have the expressions
\begin{align}
M_{fs}^{ext:fs,\mu} =  \sum_{i=1}^n \left( \prod_{j\neq i}  J^{f}(p_j) \right)  J^{fs,\mu}(p_i + q - \hk , \hk,q ) \otimes H_i^{fs}(\hp_1,...;\hp_i+q-\hk,\hk;...,\hp_n)\, ,
\end{align}
when radiating a photon from the $fs$-jet, and
\begin{align}
M_{fs}^{ext:f,\mu} =  \sum_{i=1}^n \sum_{l\neq i}  \left( \prod_{j\neq i,l}  J^{f}(p_j) \right)  J^{f,\mu}(p_l + q , q )  J^{fs}(p_i  - \hk , \hk) \otimes  H_i^{fs}(\hp_1,...,\hp_l+q,...;\hp_i-\hk,\hk;...,\hp_n)\, ,
\end{align}
when radiating from a leading jet, and finally
\begin{align}
M_{fs}^{int,\mu} =  \sum_{i=1}^n \left( \prod_{j\neq i}  J^{f}(p_j) \right)  J^{fs}(p_i  - \hk , \hk) \otimes  H_i^{fs,\mu}(\hp_1,...;\hp_i-\hk,\hk;...,\hp_n,q)\, 
\end{align}
when radiating from the hard part. This factorization of photon emission is illustrated in Fig.\ \ref{fig:low_argument}.

The three radiative amplitudes we have identified are related by the Ward identity, which takes the form
\begin{align} \label{eq:ward_identity_fs}
q_\mu M_{fs}^{ext:fs,\mu} + q_\mu M_{fs}^{ext:f,\mu}  + q_\mu M_{fs}^{int,\mu} = 0 \, .
\end{align}
We recall that this Ward identity applies to each diagram represented in Fig. \ref{fig:jet_ward_identity} individually and there is no need to consider all diagrams contributing to an amplitude at once.

The three special cases of the jet Ward identity from Fig.\ \ref{fig:jet_ward_identity} that will be of use to us can be stated as follows using our notation
\begin{align}\label{eq:jet_ward_identity}
q_\mu  J^{fs,\mu}(p_i + q -\hk , \hk,q ) &=  e_i J^{fs}(p_i - \hk, \hk) \, , \nonumber \\
q_\mu  J^{fss,\mu}(p_i + q - \hk_1 -\hk_2 , \hk_1, \hk_2, q ) &=  e_i J^{fss}(p_i - \hk_1 -\hk_2, \hk_1, \hk_2) \, , \nonumber \\
q_\mu  J^{fff,\mu}(p_i + q - \hk_1 - \hk_2 , \hk_1, \hk_2 ,q ) &  =  e_i J^{fff}(p_i - \hk_1 - \hk_2 , \hk_1, \hk_2 ) \nonumber \\
& \quad + e_i J^{fff}(p_i + q - \hk_1 - \hk_2 , \hk_1 - q, \hk_2 ) \nonumber \\
& \quad -  e_i  J^{fff}(p_i+q - \hk_1 - \hk_2 , \hk_1, \hk_2 - q ) \, ,
\end{align}
with $e_i$ the charge of the $i^{th}$ scattering fermion or antifermion. On the left hand side in each case, we choose to route the photon momentum $q$ into the hard function via the through-going fermion line that becomes the external particle. The momenta $k$,  $k_1$, and $k_2$ are collinear momenta that become part of a loop once the jet is attached to its corresponding hard part. These Ward identities apply whether the scalars, which are neutral, are soft or collinear.

Considering an $fs$-jet, the jet Ward identities allow us to expand \eqref{eq:ward_identity_fs} and cast it into the explicit form
\begin{align} \label{eq:ward_identity_fs_expanded}
 \sum_{i=1}^n & \left( \prod_{j\neq i}  J^{f}(p_j) \right)  J^{fs}(p_i  - \hk , \hk) \otimes q_\mu  H_i^{fs,\mu}(\hp_1,...;\hp_i-\hk,\hk;...,\hp_n,q) \nonumber \\
&= - \sum_{i=1}^n \left( \prod_{j\neq i}  J^{f}(p_j) \right)  e_i J^{fs}(p_i - \hk, \hk) \otimes  H_i^{fs}(\hp_1,...;\hp_i+q-\hk,\hk;...,\hp_n) \nonumber \\
& \quad -\sum_{i=1}^n \sum_{l\neq i}  \left( \prod_{j\neq i,l}  J^{f}(p_j) \right) e_l J^{f}(p_l )  J^{fs}(p_i - \hk , \hk) \otimes  H_i^{fs}(\hp_1,..., \hp_l+q,...; \hp_i-\hk, \hk;..., \hp_n) \, .
\end{align}
The next step is to Taylor expand the non radiative hard parts to $O(\lambda^2)$ in order to deduce the radiative hard part $ H_i^{fs,\mu}(\hp_1,...; \hp_i-\hk, \hk;..., \hp_n, q)$ to $O(\lambda^0)$. We will suppress any consideration of the transition from the radiative kinematics to an elastic configuration such as the one we considered in detail in Sec. \ref{sec:review_low_theorem}. Our analysis has shown that the $\xi_i$'s of Burnett and Kroll can be constructed in general and will not affect the formula we obtain for the internal amplitude. The expansion of the hard part then, is
\begin{align}\label{eq:taylor_hard_one}
 H_i^{fs}& (\hp_1,...;\hp_i+q-\hk, \hk;..., \hp_n) =\nonumber \\
&  \,  H_i^{fs}(\hp_1,...;\hp_i-\hk, \hk;..., \hp_n) + \, q^\mu \frac{\partial}{\partial \hp_i^\mu} H_i^{fs}(\hp_1,...;\hp_i-\hk, \hk;...,\hp_n) + O(\lambda^4) \, ,
\end{align}
for the photon attached to the $fs$-jet $i$, and,
\begin{align}\label{eq:taylor_hard_two}
 H_i^{fs}& (\hp_1,..., \hp_l+q, ...; \hp_i- \hk, \hk;..., \hp_n) =\nonumber \\
&  \,  H_i^{fs}(\hp_1,...; \hp_i- \hk, \hk;..., \hp_n) + \, q^\mu \frac{\partial}{\partial \hp_l^\mu} H_i^{fs}(\hp_1,..., \hp_l,...;  \hp_i- \hk, \hk;..., \hp_n) + O(\lambda^4) \, ,
\end{align}
when $l \neq i$ is a leading power jet. The derivative above acts on all the components of each $\hp_i$ for $i = 1, \dots, n$. These are treated as variables $p_i^\mu$ that the hard part depends upon for the sake of the differentiation.  Once the differentiation has been performed, the vectors $p_i^\mu$ are evaluated at the collinear configurations $\hp_i^\mu$ of the corresponding $p_i$'s.

Substituting the expansions \eqref{eq:taylor_hard_one} and \eqref{eq:taylor_hard_two}  into \eqref{eq:ward_identity_fs_expanded}, we obtain
\begin{align} \label{eq:equation_solve_hard}
 \sum_{i=1}^n & \left( \prod_{j\neq i}  J^{f}(p_j) \right)  J^{fs}(p_i  - \hk , \hk) \otimes  q_\mu  H_i^{fs,\mu}(\hp_1,...; \hp_i- \hk, \hk;..., \hp_n,q) \nonumber \\
&= - \sum_{i=1}^n  \left( \prod_{j\neq i}  J^{f}(p_j) \right)  J^{fs}(p_i - \hk, \hk)  \nonumber \\
& \quad \otimes \sum_{l=1}^{n} \left[ e_l \,  H_i^{fs}(\hp_1,...; \hp_i-\hk, \hk;..., \hp_n) +  e_l \, q^\mu \frac{\partial}{\partial \hp_l^\mu} H_i^{fs}(\hp_1,...; \hp_i-\hk, \hk;..., \hp_n)\right] \nonumber \\
& \quad + O(\lambda^4) \, .
\end{align}
After summing over the index $l$, the first term in the square brackets above vanishes by charge conservation, as in the standard Low analysis.  This cancellation confirms that contributions from the hard part associated with $fs$-jets are higher order in $\lambda$, and decouple from the use of the Ward identity for the leading jets. The natural solution to Eq.\ \eqref{eq:equation_solve_hard} for the radiative hard part is
\begin{align}\label{eq:hard_particular_solution}
 H_i^{fs,\mu}(\hp_1,...; \hp_i-\hk, \hk;..., \hp_n,q) =  -\sum_{l=1}^{n}  e_l \,  \frac{\partial}{\partial \hp_{l\mu}} H_i^{fs}( \hp_1,...; \hp_i- \hk, \hk;..., \hp_n) + O(\lambda^2) \,.
\end{align}
In fact, \eqref{eq:hard_particular_solution} is the full solution to Eq. \eqref{eq:equation_solve_hard}. Additional gauge invariant terms would require the radiative hard part to have enhancements in $\lambda$ -- see the discussion of Eq. \eqref{eq:gauge_invariant_terms}. These are not allowed since any dependence of the hard part on $q$ or $m$ comes as a subleading correction to the hard exchanges.

Since $f\partial s$-jets correspond to higher order contributions to $fs$-jet amplitudes, they are treated using the same steps. Similarly, the analysis for photon emission from diagrams with a single $fss$-jet is virtually unchanged. We need only include two collinear scalar momenta $k_1$ and $k_2$ rather than a single one. The final result is
\begin{align}
 H_i^{fss,\mu}(\hp_1,...; & \hp_i- \hk_1- \hk_2,  \hk_1, \hk_2;..., \hp_n,q) \nonumber \\
& =  -\sum_{l=1}^{n}  e_l \,  \frac{\partial}{\partial \hp_{l\mu}} H_i^{fss}(\hp_1,...; \hp_i- \hk_1- \hk_2, \hk_1, \hk_2;..., \hp_n)  + O(\lambda^2)\,.
\end{align}

For diagrams with $fff$-jets, the analysis involves one additional step because the corresponding jet Ward identity has three terms as shown in Eq.\ \eqref{eq:jet_ward_identity} and Fig.\ \ref{fig:jet_ward_identity}. The momentum flow through the jets enables us to shift the collinear loop integration momenta, after which we obtain
\begin{align}
H_i^{fff,\mu}&(\hp_1,...; \hp_1 - \hk_1 - \hk_2, \hk_1, \hk_2; ..., \hp_n, q) = \nonumber \\
&- \sum_{l=1}^n e_l \frac{\partial}{\partial \hp_{l\mu}} H_i^{fff} ( \hp_1, ...; \hp_i - \hk_1 - \hk_2, \hk_1, \hk_2; ..., \hp_n) \nonumber \\
&- e_i \frac{\partial }{ \partial \hk_{1\mu}} H_i^{fff} (\hp_1, ...; \hp_i - \hk_1 - \hk_2 , \hk_1, \hk_2; ..., \hp_n) \nonumber \\
&+ e_i  \frac{\partial }{ \partial \hk_{2\mu}} H_i^{fff} (\hp_1, ...; \hp_i - \hk_1 - \hk_2 , \hk_1, \hk_2; ..., \hp_n) + O(\lambda^2) \, .
\end{align}
In the above, the partial derivatives $\frac{\partial}{\partial \hk_{l\mu}}$ only act on the explicit dependence of $H_i^{fff}$ on $\hk_1$ and $\hk_2$. That is, there is no contribution to the derivative from the implicit dependence of $H_i^{fff}$ on $\hk_1$ and $\hk_2$ through $\hp_i - \hk_1 - \hk_2$.

This concludes our description of photon emission at orders beyond $O(\lambda^0)$. In the next section, we return to Eq.\ \eqref{eq:preliminary_low} and analyze the external amplitude using the $KG$ decomposition.




\section{The $KG$ decomposition} \label{sec:kg_decomposition}

By making use of the jet Ward identities once again, it is possible to unravel some structure in the small $q$ expansion of the radiative jet functions. Following del Duca \cite{DelDuca:1990gz}, who drew inspiration from Grammer and Yennie \cite{Grammer:1973db}, we consider the two tensors
\begin{align} \label{eq:KG_definition}
\tensor{K}{_{i\,}_\mu^\nu} &\equiv \frac{(2p_i + q)_\mu \, q^\nu }{2 p_i\cdot q + q^2} \, , \nonumber \\
\tensor{G}{_{i\,}_\mu^\nu} &\equiv \tensor{g}{_\mu^\nu} - \tensor{K}{_{i\,}_\mu^\nu} \, .
\end{align}
We will use the $K$ and $G$ tensors to decompose the soft photon polarization $\epsilon^\mu(q)$  into two complementary polarizations. It will turn out that the $K$ polarized photon emission amplitude contains all the leading $O(\lambda^{-2})$ terms while the $G$ polarized photon amplitude supplies transverse corrections that begin at $O(\lambda^{-1})$. Note that transversality of $G$ polarized photons, $q^\mu \, \tensor{G}{_{i\,}_\mu^\nu} = 0$, follows immediately from \eqref{eq:KG_definition}.

 So far, we have only been considering the stripped amplitude $M_\mu$, that is, we have derived the photon emission amplitude with the photon polarization tensor $\epsilon^\mu(q)$ stripped away. For the purposes of applying the $KG$ decomposition, it is useful to reintroduce this polarization tensor. 

Consider first the emission of a $K$ polarized photon. For definiteness, we will illustrate our argument using $fss$-jets, although the same conclusion applies to any type of jet. The relevant identity is
\begin{align}
\epsilon^\mu(q) & \tensor{K}{_{i\,}_\mu^\nu} J_{i,\nu}^{fss}(p_i+q-\hk_1-\hk_2, \hk_1, \hk_2) \otimes H_i^{fss}(\hp_i+q-\hk_1-\hk_2, \hk_1, \hk_2) \nonumber \\
&= e_i \, \frac{\epsilon \cdot (2 p_i + q)}{2 p_i \cdot q + q^2} \, J_i^{fss}(p_i-\hk_1-\hk_2, \hk_1, \hk_2) \otimes  H_i^{fss}(\hp_i+q-\hk_1-\hk_2, \hk_1, \hk_2) \, .
\end{align}
This result follows immediately from the application of the jet Ward identity for $fss$-jets, as shown in \eqref{eq:jet_ward_identity}. Since the infrared degree of divergence of a non radiative $fss$-jet is $\gamma = 2$, the above formula confirms that the emission of a $K$ polarized photon starts at $O(\lambda^0)$. The same conclusion holds when attaching the soft photon to an $f\partial s$-jet, an $fff$-jet, or to any jet in diagrams containing either two $fs$ -jets or two $f \softs$-jets as these all have degree of divergence $\gamma = 2$ prior to the soft photon insertion. For a diagram with a single $fs$-jet, the emission of a $K$ polarized photon is $O(\lambda^{-1})$ since $fs$-jets have degree of divergence $\gamma =1$ in the massive case. Finally, following the same reasoning, the emission of a $K$ photon from a leading jet is $O(\lambda^{-2})$. Therefore, the $K$ polarization tensor does contain a leading order term, which is derived purely from the application of the jet Ward identity. The question is then whether a leading term also appears in the complimentary polarization.

A $G$ polarized photon is connected to the radiative jet function through the insertion of a field strength tensor operator,
\begin{align}
\epsilon^\mu(q) \, \tensor{G}{_{i\,}_\mu^\nu} = \frac{(2p_i+q)_\mu}{2 p_i \cdot q + q^2} F^{\mu\nu}(q, \epsilon) \, ,
\end{align}
where $F^{\mu\nu}(q,\epsilon) = q^\mu \epsilon^\nu (q) - q^\nu \epsilon^\mu(q)$.  An important property of the $G$ polarization tensor following from this form is that it annihilates the scalar photon vertex,
\begin{align}
\epsilon^\mu (q)\, \tensor{G}{_{i\,}_\mu^\nu} (2 p_i + q )_\nu = 0 \, .
\end{align}
In particular, this implies that $ \epsilon^\mu (q)\,  \tensor{G}{_{i\,}_\mu^\nu} p_{i\nu} = O(q)$. An analysis of the general loop integrand for jet functions shows that their leading term is always proportional to $p_{i\nu}$, where $p_i$ is the external momentum of the jet. The first subleading term is suppressed by at least one power of $\lambda$ in the massive fermion case. Since contracting $p_{i\nu}$ with the $G$ polarization tensor yields a suppression of $\lambda^2$, we find that the emission of a $G$ polarized photon is suppressed by at least one power of $\lambda$ relative to the corresponding $K$ polarized emission \cite{DelDuca:1990gz}.

Using the techniques of Sec. \ref{sec:power_counting_applied}, we found that attaching a soft photon to a non radiative diagram with an $f\partial s$,  $fss$, or $fff$-jet, as well as to a diagram with two non radiative $fs$ or $f\softs$-jets makes the diagram at most logarithmic in $q$. Hence, emission of a $G$ photon starts at $O(\lambda)$ when attaching a soft photon to any of those diagrams, which is beyond the order of accurary of our extension of Low's theorem to high energies. When a diagram only has a single $fs$-jet, the emission of a $G$ photon will start at $O(\lambda^{0})$. Finally, for leading jets, $G$ photon emission begins at $O(\lambda^{-1})$ when considering massive fermions.

We mention that using an on-shell renormalization scheme further simplifies external emission amplitudes. It is straightforward to verify that in the tree level radiative leading jet, $G$ photon emission starts at $O(\lambda^0)$, and the leading $O(\lambda^{-2})$ term is entirely contained within the $K$ photon emission amplitude. By definition, an on-shell scheme eliminates the $q\rightarrow 0$ limit of the radiative loop diagrams. Consequently, all leading $O(\lambda^{-2})$ behavior is contained in the tree level diagram and fully accounted for by $K$ photon emission. Further, $G$ photon emission begins at $O(\lambda^0)$ in this scheme.

We now have all the required pieces to apply the $KG$ decomposition to all external radiative terms in \eqref{eq:preliminary_low} and thereby complete the derivation of the final form of our extension of Low's theorem. It is useful to separate the soft photon amplitude into three contributions: the internal emission amplitude, the external emission amplitude for $K$ polarized photons, and the external emission amplitude for $G$ polarized photons,
\begin{align} \label{eq:final_low}
\epsilon \cdot M &= \epsilon \cdot M_{int}  + \epsilon \cdot M_{ext}^K + \epsilon \cdot M_{ext}^G  \, .
\end{align}

The internal emission amplitude simply follows from contracting the photon polarization tensor with the first term in \eqref{eq:preliminary_low},
\begin{align}\label{eq:int_photon}
\epsilon \cdot M_{int} &= - \left(\prod_{i=1}^n J_i^f \right)\otimes \sum_{l=1}^n e_l \, \epsilon^\mu \frac{\partial}{\partial \hp_l^\mu} H \bigg|_{P_0} + O(\lambda) \, ,
\end{align}
where as above, $P_0$ indicates that we are evaluating the derivative at a set of momenta constructed from the procedure described in Sec.\ \ref{sec:review_low_theorem}.

The $KG$ decomposition allows us to extract the leading $O(\lambda^{-2})$ term from the radiative jet functions. This leading term is contained within the complete $K$ polarized emission amplitude, which is
\begin{align} \label{eq:ext_k_photon}
\epsilon \cdot M_{ext}^K  &= \sum_{i=1}^n \left( \prod_{j=1}^n J_j^f \right) \otimes e_i \frac{\epsilon \cdot (2 p_i +q)}{2 p_i \cdot q + q^2}\, H(\hp_i + q) \nonumber \\
&+ \sum_{i=1}^n \sum_{\theta \in \Theta_1} \left( \prod_{j \neq i} J_j^f \right) J_i^\theta \otimes \sum_{h=1}^n e_h \frac{ \epsilon \cdot (2 p_h + q)}{2 p_h \cdot q + q^2}\, H_i^\theta(  \hp_h + q) \nonumber \\
&+\frac{1}{2} \sum_{i \neq j} \sum_{\theta \in \Theta_2} \left( \prod_{l \neq i,j} J_l^f \right) J_i^\theta J_j^\theta S^\theta \otimes \sum_{h=1}^n e_h \frac{ \epsilon \cdot (2 p_h + q)}{2 p_h \cdot q + q^2}\, H_{ij}^{\theta\theta}(\hp_h + q) \nonumber \\
&+ O(\lambda) \, ,
\end{align}
where again $\Theta_1 = \{ fs, \, f\partial s, \, fss, \, fff \}$ and $\Theta_2 = \{ fs, \, f\softs \}$, and $S^\theta = 1$ if $\theta = fs$ and  $S^\theta = S$ if $\theta = f\softs$. In each term, the soft photon polarization is coupled to a tree level leading factor reminiscent of early treatments of the soft theorem  \cite{Low:1958sn,Burnett:1967km,Weinberg:1965nx}. We have indicated which argument of the hard parts is shifted by $q$. One could extend the construction of the $\xi_i$'s and $\eta_i$'s of Sec.\ \ref{sec:review_low_theorem} and expand the hard parts. However, corrections of order $O(\lambda^2)$ would only be required for the hard part corresponding to the leading jets, $H(\dots, \hp_i + q, \dots)$.

 Corrections to the soft theorem also appear as separately transverse emission amplitudes which couple to a tree level leading order factor through the field strength tensor in momentum space. These are always suppressed by at least one power of $\lambda$ relative to the corresponding $K$ polarized photon emission amplitude. Therefore, at our order in $\lambda$, only leading and $fs$-jets are relevant for these corrections, which are given by
\begin{align} \label{eq:ext_g_photon}
\epsilon \cdot M_{ext}^G &= \sum_{i=1}^n \sum_{l\neq i } \left( \prod_{j \neq i, l} J_j^f \right) \frac{(2p_l + q)_\mu }{2 p_l \cdot q + q^2} F^{\mu\nu}(q,\epsilon) \, J_{l,\nu}^f J_i^{fs} \otimes H_i^{fs} (\hp_l + q) \nonumber \\
&\quad + \sum_{i=1}^n \left( \prod_{j \neq i} J_j^f \right) \frac{(2 p_i + q )_\mu }{2 p_i \cdot q + q^2} F^{\mu\nu}(q, \epsilon) \left( J_{i,\nu}^f + J_{i,\nu}^{fs}   \right) \otimes H_i^{fs} ( \hp_i + q) \nonumber \\
&\quad + O(\lambda) \, .
\end{align}
Equations \eqref{eq:final_low}-\eqref{eq:ext_g_photon} taken together make up our final version of Low's theorem.




\section{Conclusion} \label{sec:conclusion}

Inspired by the renewed interest in soft theorems, we have set out to investigate the role of loop corrections at high energy in Low's classic result. Work on this subject had already been carried out by del Duca \cite{DelDuca:1990gz} who showed that in the limit of high center of mass energy $E$, Low's argument only applies in the vanishingly small region $q \ll m^2/E$. In the regime where $q \sim m^2/ E$, del Duca identified loop corrections that take the form of universal infrared sensitive matrix elements, the jet functions.  To identify these loop corrections, he needed to adapt Low's analysis by factorizing the elastic amplitude into jet functions and a hard part, and then to consider separately photon emission from each factor. More recently, an analysis of soft theorems in effective field theory has been given in \cite{Larkoski:2014bxa}.

Focusing on Yukawa and scalar theories, we revisited the main assumption underlying del Duca's treatment of the soft photon theorem, namely the factorized form of the elastic amplitude to which a soft photon is attached to construct the radiative amplitude  for massive and massless fermions. The application of the techniques described here to QED and other gauge theories \cite{Bonocore:2015esa} is left for future work. 

The first step in deriving a factorized amplitude is to identify the regions of loop space giving rise to non analytic terms in the full loop integrals. The effect of the singular regions of loop momentum space is then reproduced by universal functions that have the same singularity structure as the original amplitude. The non-singular parts of loop space are accounted for by a set of hard functions.  As we have reviewed in Sec. \ref{sec:power_counting}, singular regions of loop space are classified according to their degree of divergence $\gamma$, which indicates an overall scaling of $\lambda^\gamma$ with $\lambda \equiv m/E$. In particular, del Duca's analysis considers attaching a soft photon to the pinch surface with minimal $\gamma$. However, in the region $q \sim \lambda^2 E$, the soft photon theorem is an expansion going from order $\lambda^{-2}$ to $\lambda^{0}$. Since attaching a soft photon to a collinear fermion line reduces the degree of divergence of a non radiative diagram by $2$, it is clear that to obtain all contributions to the radiative amplitude up to $O(\lambda^0)$, one needs to attach the soft photons to non radiative diagrams with a scaling up to $\lambda^2$. Del Duca was concerned with QED, and therefore, we may not directly compare our results with his. However, it is natural to suspect that, compared to Ref.\ \cite{DelDuca:1990gz}, additional terms may occur from matrix elements with $\gamma=-1$ and $\gamma = 0$  in certain amplitudes in gauge as well as Yukawa theories \cite{Larkoski:2014bxa}.

The new terms originate from our non leading jets: the $fs$, $fss$, and $fff$-jets. As we have seen, there are also contributions from diagrams with soft two-point functions, which is qualitatively new. The reduced diagrams corresponding to the new sources of terms in the soft theorem were shown in Figs. \ref{fig:contribution_low_analysis} and \ref{fig:contribution_massive}. We emphasize that our treatment takes into account all infrared sensitive behavior of the radiative amplitude at all loop orders in the region $q \sim m^2/E$. In particular, we do not restrict ourselves to the massless case, although our results are easily adapted to this limit.  The full list of contributions to the soft theorem from non leading jets is generated by attaching a photon to the diagrams of Figs. \ref{fig:contribution_low_analysis} and \ref{fig:contribution_massive}. Factorizing the radiative amplitude into emission from all components identified in such a list gives rise to the formula shown in Eq. \eqref{eq:generic_radiative}.

Having derived the proper factorization of the radiative amplitude, we obtained the final form of our extension of Low's theorem by applying del Duca's technique. This involves the application of the jet Ward identities \eqref{eq:jet_ward_identity} to derive the radiative hard part, followed by an application of the $KG$ decomposition to isolate the leading term from subleading corrections. The end result of this procedure is given in formulas \eqref{eq:final_low}, \eqref{eq:int_photon}, \eqref{eq:ext_k_photon}, and \eqref{eq:ext_g_photon}. Of course, the leading term retains its form as in Low's classic result. Loop corrections from both leading and non leading jets significantly alter the subleading term, however. Further, the jet functions give rise to separately transverse contributions to the external amplitude, as shown in \eqref{eq:ext_g_photon}.

Our results in Yukawa and scalar theories are interesting in their own right because of their potential applications to non linear sigma models and pion scattering. However, they can also be viewed as a testing ground for gauge theories. One theory of particular interest to us is of course QCD. Work in this direction has already been undertaken in Refs. \cite{Larkoski:2014bxa,Bonocore:2015esa, Bonocore:2016awd}. 

A related subject concerns the emission of soft gravitons from Yukawa and scalar theories. The power counting analysis in this case proves to be more challenging because the soft graviton can couple to both fermions and scalars as well as to the potential itself. This is the subject of ongoing work.

\section*{Acknowledgements}
The author is grateful to George Sterman for suggesting this problem and for many helpful discussions and suggestions. The author also thanks the Fonds de recherche du Qu\'ebec --  nature et technologies for their financial support through their doctoral research scholarships. This work was supported in part by the National Science Foundation, grants
PHY-1316617 and 1620628.

\bibliography{bib_soft_theorem_yukawa_scalar}{}
\bibliographystyle{ieeetr}

\end{document}